\documentclass[a4paper,fleqn]{cas-dc}

\usepackage[numbers,sort&compress]{natbib}

\usepackage{amsmath}
\usepackage{amssymb}
\usepackage{algorithm}
\usepackage{algorithmicx}
\usepackage{algpseudocode}
\usepackage{color}
\usepackage{multirow}
\usepackage{makecell}
\usepackage{soul}

\usepackage{threeparttable}
\usepackage{subfig}

\newtheorem{remark}{Remark}

\def\tsc#1{\csdef{#1}{\textsc{\lowercase{#1}}\xspace}}
\tsc{WGM}
\tsc{QE}

\begin{document}
\let\WriteBookmarks\relax
\def\floatpagepagefraction{1}
\def\textpagefraction{.001}

\shorttitle{Robust Deep Reinforcement Learning for Inverter-based Volt-Var Control in  \textcolor{blue}{Partially Observable} Distribution Networks}    

\shortauthors{Q. Liu et al.}

\title [mode = title]{Robust Deep Reinforcement Learning for Inverter-based Volt-Var Control in  \textcolor{blue}{Partially Observable} Distribution Networks}


\tnotetext[1]{This work was supported in part by the National Key R\&D Program of China (2020YFB0906000, 2020YFB0906005).}

%

\author[1]{Qiong Liu}[orcid=0000-0002-5244-7096]



\ead{liuqiong_yl@outlook.com}


\credit{Conceptualization of this study, Methodology, Writing, Software}

\author[1]{Ye Guo}[orcid=0000-0002-5268-5289]

\cormark[1]

\ead{guo-ye@sz.tsinghua.edu.cn}


\credit{Conceptualization of this study, Writing-review \& editing, Supervision}

\author[2]{Tong Xu}[orcid=0000-0002-7432-808X]


\ead{taterx@foxmail.com}


\credit{Conceptualization of this study, Writing-review \& editing}

\affiliation[1]{organization={ Tsinghua-Berkeley Shenzhen Institute, Tsinghua University},
            city={Shenzhen},
            postcode={518071}, 
            state={Guangdong},
            country={China}}

\affiliation[2]{organization={Southwest Electric Power Design Institute Co., Ltd. of China Power Engineering Consulting Group},
            city={Chengdu},
            postcode={},
            state={Sichuan},
            country={China}}

\cortext[1]{Corresponding author: Ye Guo}



\begin{abstract}
Inverter-based volt-var control is studied in this paper. One key issue in DRL-based approaches is the limited measurement deployment in active distribution networks, which leads to problems of a partially observable state and unknown reward.
To address those problems, this paper proposes a robust DRL approach with a conservative critic and a surrogate reward.
The conservative critic utilizes the quantile regression technology to estimate conservative state-action value function based on the partially observable state, which helps to train a robust policy;
the surrogate rewards of power loss and voltage violation are designed
that can be calculated from the limited measurements.
The proposed approach optimizes the power loss of the whole network and the voltage profile of buses with measurable voltages while indirectly improving the voltage profile of other buses.
Extensive simulations verify the effectiveness of the robust DRL approach in different limited measurement conditions, even when only the active power injection of the root bus and less than $10\%$ of bus voltages are measurable.
\end{abstract}


\maketitle










\section{Introduction}

To achieve a carbon-neutral society, active distribution networks (ADNs) will incorporate more renewable distributed generators (DGs) like PV and wind turbines. The generation output of these DGs is highly volatile and uncertain, which can cause problems such as voltage violations and increased power loss. Meanwhile, most DGs are inverter-based (IB) devices that are capable of fast reactive control.
This makes it increasingly attractive to use these IB devices in real-time volt-var control (VVC), which minimizes power loss and optimizes voltage profiles in ADNs \cite{farivarInverterVARControl2011, chenPhysicalassistedMultiagentGraph2023}. Such a voltage control paradigm is also known as ``IB-VVC" \cite{juBiLevelConsensusADMMBased2022}.

However, for ADNs with massive inverter-based energy resources, three challenges may emerge in VVC algorithm design:
1) IB-VVC needs to make real-time decisions to mitigate voltage fluctuations due to the high volatility and uncertainty of renewable energy\cite{yanMultiAgentSafeGraph2024}.
2) Current models of ADNs usually involve significant errors or even unknown parts \cite{wangSafeOffPolicyDeep2020}.
3) ADNs usually have limited measurement deployments \cite{xuDatadrivenInverterbasedVolt2023}.


Model-based IB-VVC approaches have severe limitations in addressing those challenges. It is a nonlinear programming problem requiring time-consuming iterative steps to converge for off-the-shelf NLP solvers such as IPOPT solvers \cite{wangSafeOffPolicyDeep2020}.
{\color{blue}

}

The performance of the model-based approach is vulnerable to parameter errors or unknown model \cite{caoPhysicsInformedGraphicalRepresentationEnabled2024}.
In addition, when some load and generation values are unknown, model-based approaches are difficult to implement. One of the methods is to supplement those unknown data using pseudo-measurements, but those data are not real-time and may be full of noise \cite{azimianStateTopologyEstimation2022}.
State estimation can estimate the unknown data and filter out noise, which has been widely investigated \cite{primadiantoReviewDistributionSystem2017}. However, it relies heavily on the accurate power flow model and may not work well for ADNs with inaccurate or unknown models.
Sometimes, the pseudo-measurements may not be available for ADNs with weak measurement conditions.

Existing literature has fully demonstrated the advantages of DRL approaches in addressing the former two challenges \cite{liuTwoStageDeepReinforcement2021, liuRobustRegionalCoordination2021, gaoModelaugmentedSafeReinforcement2022, caoDeepReinforcementLearning2022, 
caoMultiAgentDeepReinforcement2020,
yiRealTimeSequentialSecurityConstrained2024, 
sunOptimalVoltVar2024, peiMultiTaskReinforcementLearning2023, chenReinforcementLearningSelective2022}.
First, in the application stage, DRL achieves real-time control by performing a fast-forward calculation of policy networks. The time-consuming optimization process is shifted to the training stage \cite{yiRealTimeSequentialSecurityConstrained2024, caoMultiAgentDeepReinforcement2020}.
Second, DRL is a model-free method that learns to make near-optimal decisions by interacting with the ADN environments \cite{chenReinforcementLearningSelective2022}.
However, there is still insufficient research on robust DRL algorithms that are applicable to limited measurement conditions.

DRL is a potential method to address the challenge of limited measurement deployments, but it has seldom been studied.
The limited measurements lead to the problem of a partially observable state and unknown reward of the Markov decision process (MDP) in DRL.
In cases where the reward is known, some DRL algorithms, such as SAC \cite{haarnojaSoftActorCriticAlgorithms2019} and PPO \cite{schulmanProximalPolicyOptimization2017}, can be applied directly to the partially observable tasks with slight modifications \cite{mengPartialObservabilityDRL2022}.
However, the decision performance may be affected due to the missing real-time measurements \cite{yanMultiAgentSafeGraph2024}.
For IB-VVC tasks, we can also use pseudo-measurements to complement the unknown state and introduce a physics-informed global graph attention network and a deep auto-encoder to filter the noise of the pseudo-measurements \cite{caoPhysicsInformedGraphicalRepresentationEnabled2024}.
The work follows the route of state estimation that introduces the pseudo measurements to improve the observation level, which may be not suitable for ADNs with only limited real-time measurements.
Additionally, the unknown reward led by limited real-time measurements has not been studied.

The partially observable state problems also exist in multi-agent DRL for decentralized VVC tasks, but they are completely different from the problems studied in this paper.
For multi-agent DRL, the partially observable state only occurs in actors, and critics can still observe the global state \cite{liuOnlineMultiAgentReinforcement2021, liuRobustRegionalCoordination2021, sunTwoStageVoltVar2021, yanMultiAgentSafeGraph2024}.
This is because decentralized VVC assumes sufficient measurement deployment but has delayed uploading of measured information due to communication limitations.

Given the literature above, 
we observe two gaps in studying the problem of limited measurement deployment of ADNs for DRL-based VVC tasks:
1) Paper \cite{caoPhysicsInformedGraphicalRepresentationEnabled2024} introduces pseudo-measurements to address the problem of partially observable states. However, sometimes, those pseudo-measurements are difficult to acquire in ADNs. Further research is required for DRL to address the problem.
2) The limited measurement deployments also lead to the unknown reward, which has not been studied.

\begin{figure}[!ht]
\centering
\includegraphics[width=2.4in]{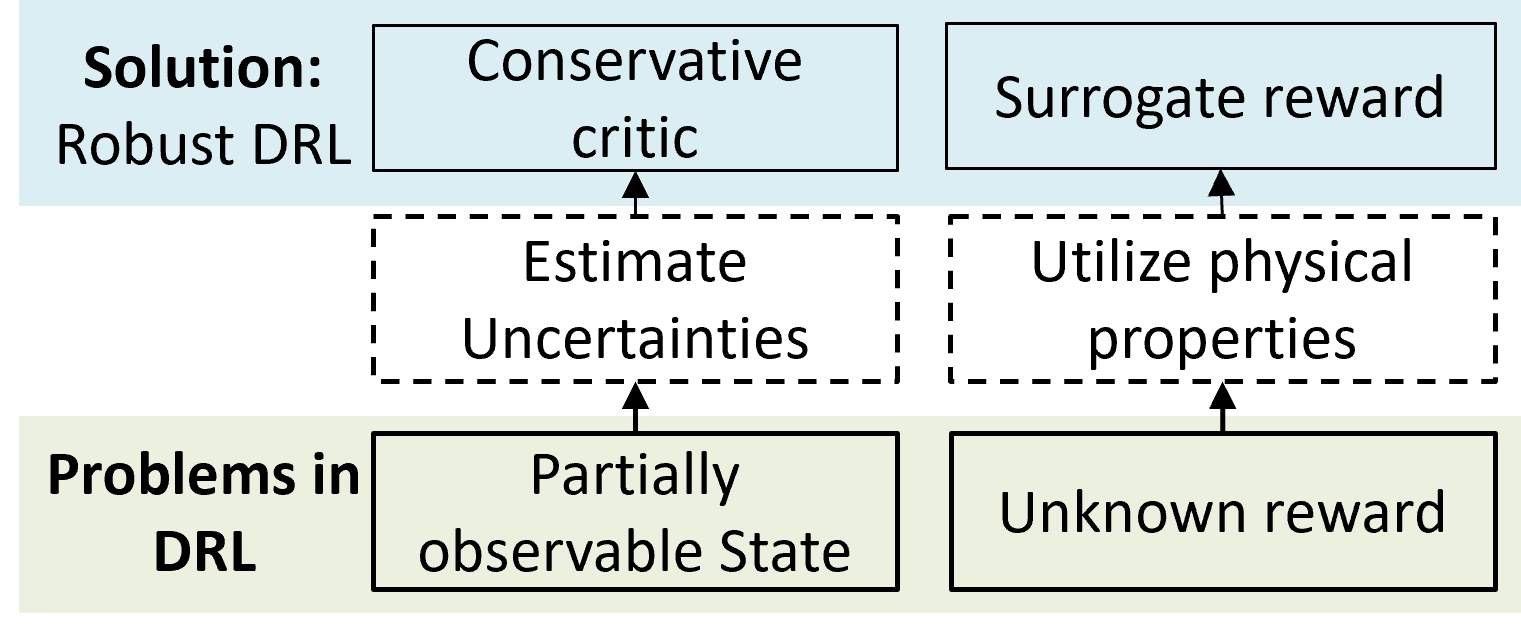}
\caption{The overview of the proposed robust deep reinforcement learning approach.
}\label{overview_robustDRL}
\end{figure}

To address the problems of the partially observable state and the unknown reward, this paper proposes a robust DRL approach, illustrated in Fig. \ref{overview_robustDRL}.
For the problem of the partially observable state, we utilize the conservative critic based on the quantile regression technology to estimate the conservative state-action value function, which helps to train a robust policy.
For the problem of the unknown reward, we leverage the two physical knowledge of VVC tasks to design the surrogate reward of power loss and voltage violation.
\textit{Physical knowledge 1:}
For IB-VVC tasks, minimizing the power loss of the entire network is equivalent to minimizing the active power injection of the root node.
Therefore, we can design the indirect reward of power loss as the active power injection of the root node.
To reflect the power loss change explicitly, we design the surrogate reward, which is the change of the indirect reward of power loss before and after the action execution.
\textit{Physical knowledge 2:} In an ADN, the voltages of neighbor buses are coupled. This means that if we optimize the measurable bus voltages, the unmeasurable ones will also be optimized accordingly.
Therefore, the surrogate reward of voltage violation is designed as the sum of the voltage violation rate of the measurable node.
The proposed robust DRL approach works for ADNs with different limited measurement deployments.
The minimal measurement condition is that the active power of the root bus and the bus voltages that need to be optimized are measured in real-time. 
{\color{blue}
We utilize ``robust" to name the proposed robust approach because it addresses the uncertainties led by partial observable states, which is similar to other robust optimization methods addressing the problems of disturbances such as measurement errors or system uncertainties.
}
Compared with the existing DRL-based VVC algorithms, the main technical advancements are summarized as follows:

\begin{itemize}

\item[1.] A conservative critic {\color{blue}is proposed} to estimate the uncertainty of the state-action value function to the partially observable states based on quantile regression technology. Then, a robust policy {\color{blue}is proposed} by maximizing the value of the conservative critic.
{\color{blue} Compared with paper \cite{liuTwoStageDeepReinforcement2021, liuRobustRegionalCoordination2021, gaoModelaugmentedSafeReinforcement2022, caoDeepReinforcementLearning2022, 
caoMultiAgentDeepReinforcement2020,
yiRealTimeSequentialSecurityConstrained2024, 
sunOptimalVoltVar2024, peiMultiTaskReinforcementLearning2023, chenReinforcementLearningSelective2022} with the standard critic, the proposed approach with the conservative critic has better performance in reducing voltage violation.}

\item[2.]
The surrogate rewards for power loss and voltage violation are proposed when the real rewards are unknown, {\color{blue} which are calculated based on the active power injection of the root bus and the measurable voltage magnitudes.}
DRL approach with the two surrogate rewards optimizes the power loss of the whole network and the voltage profile of buses with measurable voltage magnitudes, while indirectly improving the voltage profile of other buses.



\item[3.] The minimal measurement condition of the proposed robust DRL approach is identified, which is the measurement of the active power injection of the root bus and a few bus voltage magnitudes that need to be optimized.
{\color{blue} There may be no existing work that has studied the DRL-based VVC approach for ADNs with such limited measurement conditions.
}
\end{itemize}

The rest of this paper is organized as follows: 
Section II introduces the preliminaries of IB-VVC and DRL.
Section III formulates the IB-VVC as a partially observable Markov decision process (POMDP) with two extra items, the partially observable state and the surrogate reward, compared to the existing MDP.
Section IV proposes a robust DRL with the conservative critic and surrogate reward to address the two problems. Simulation and conclusions are provided in Section V and Section VI.

\section{Preliminary}




\subsection{Inverter-based Volt-Var Control}\label{mb_IB_VVC}

We considers an ADN with $N+1$ buses, which can be represented by a graph $\mathcal{G}(\mathcal{N}, \mathcal{E})$ with the collection of all buses $\mathcal{N} = 0,...,N$ and the collection of all branches $\mathcal{E} =(i, j) \in \mathcal{N}\times \mathcal{N}$. Bus $0$ denotes the graph root, and all tree branches are in the direction away from the root.


IB-VVC optimizes power loss and voltage profiles by controlling the reactive power output of distributed generators (DGs) and static Var compensators (SVCs).
It can be formulated as an optimal power flow as follows \cite{farivarInverterVARControl2011, zhangHierarchicallyCoordinatedVoltageVAR2020}:


1) The objective function of the optimization problem is minimizing the active power loss $P_{loss}$ of the networks:
\begin{equation}\label{MBO}
\begin{split}
& \min P_{loss} = \sum_{i \in \mathcal{N}} P_{Gi}  - \sum_{i \in \mathcal{N}}  P_{Di} + P_0, \\
\end{split}
\end{equation}
where 
$P_{Gi}, P_{Di}$ are the active powers of DG and load at bus $i$. 
For the buses without DG, load, or SVC, the corresponding active and reactive powers are zero. 
$P_0$ is the active power injection of the root bus.


2) The Dist-flow constraints are: 

\begin{equation}\label{EC}
\begin{aligned}
& P_j =\sum_{k: j \rightarrow k} P_{j k} - \sum_{i: i \rightarrow j}\left(P_{i j}-\frac{\left(P_{i j}\right)^2+\left(Q_{i j}\right)^2}{\left(V_i\right)^2} r_{i j}\right), \\
& Q_j =\sum_{k: j \rightarrow k} Q_{j k} -\sum_{i: i \rightarrow j}\left(Q_{i j}-\frac{\left(P_{i j}\right)^2+\left(Q_{i j}\right)^2}{\left(V_i\right)^2} x_{i j}\right) , \\
& V_j^2=V_i^2-2\left(r_{i j} P_{i j}+x_{i j} Q_{i j}\right)+\left(r_{i j}^2+x_{i j}^2\right) \frac{P_{i j}^2+Q_{i j}^2}{V_i^2},
\end{aligned}
\end{equation}
and 
\begin{equation}
\begin{aligned}
& P_j  = \begin{cases}
P_{0}, &  \ j=0,\\
P_{G j}-P_{D j}, & j = 1, \dots, N
\end{cases} \\
& Q_j = 
\begin{cases}
Q_{0}, & \  j=0,\\
Q_{G j}+  Q_{C j} -Q_{D j}, & j = 1, \dots, N,
\end{cases}  \\
\end{aligned}
\end{equation}
where 
$\forall j \in \mathcal{N}$, $P_j$, $Q_j$ denote the active and reactive power injections at bus $j$.
$P_{jk}, Q_{jk}$ denotes the active and reactive power flow from bus $j$ to $k$.
$V_i$ is the voltage magnitude at bus $i$.
$r_{ij}$, $x_{ij}$ are the resistance and reactance of branch $ij$.
$i \rightarrow j$ denotes a branch from bus $i$ to bus $j$.
$P_{Gj}$, $Q_{Gj}$ denote the active and reactive power generations of the DG at bus $j$.
$P_{Dj}$, $Q_{Dj}$ denote the active and reactive power loads at bus $j$.
$Q_{Cj}$ denotes the reactive power generation of the SVC at bus $j$.
This paper sets the reference voltage bus as $V_0 = 1  \ p.u.$ at bus 0.

3) The voltage constraints are:
\begin{equation}\label{SC}
\underline{V}_i \leq V_i \leq \bar{V}_i,
\end{equation}
where $\bar{V}_i$ and $\underline{V}_i$ are the upper and lower bounds of the voltage magnitude at bus $i$, respectively.

4) The reactive power capability constraints of DGs and SVCs are:
\begin{equation}\label{OC}
\begin{split}
\underline{Q_{Gi}} & \leq Q_{Gi} \leq \overline{Q_{Gi}},\\
\underline{Q_{Ci}} & \leq Q_{Ci} \leq \overline{Q_{Ci}},
\end{split}
\end{equation}
where 
$\overline{Q_{Gi}}, \underline{Q_{Gi}}$ and $\overline{Q_{Ci}}, \underline{Q_{Ci}}$ the upper and lower limits of reactive power generation of the DG and SVC at bus $i$, respectively.

The variables of IB-VVC can be classified as follows: $P_{D}, Q_{D}, P_{G}$ are uncontrollable variables, $V, P_0, Q_0, I$ are dependence variables, $Q_G, Q_C$ are controllable (independent) variables \cite{zhangAdvancedElectricPower2010}. Decision variables include controllable variables and dependence variables, respectively.
Giving $P_D, Q_D, P_G$, IB-VVC minimizes the power loss and ensures the voltage is in its normal range by deciding controllable variables. The dependence variables are solved as well.

\subsection{Preliminary of Deep Reinforcement Learning}


The objective of the RL algorithm is to learn a policy $\pi$ to make decisions that maximize the expected discounted accumulated reward from a sequence of MDP data $\mathcal{T}$:
\begin{equation}
    \pi^*=\arg \max _\pi \mathbb{E} [R(\mathcal{T})],
\end{equation}
where the policy is
\begin{equation}
    a \sim \pi\left(\cdot \mid s\right),
\end{equation}
the discounted accumulated reward, also named return, is
\begin{equation}
    R = \sum_{t=0}^{\infty} \gamma^t r_t .
\end{equation} the sequence of MDP data  $\mathcal{T} =(s_0,a_0,s_1,a_1, \dots )$ and, {\color{blue}$\mathbb{E}$ is a mathematical operation of expect expected value}.

For the sake of designing RL algorithms, the state-action value functions are introduced to evaluate the policy $\pi$, which is the expected return of an RL policy by starting in the state-action pair:
\begin{equation}
    Q^\pi(s, a)=\underset{a \sim \pi}{\mathbb{E}}\left[\sum_{t=0}^{\infty} \gamma^t r_t \mid s_0=s, a_0=a\right].
\end{equation}

Considering the control capabilities of inverter-based devices are fast and independent among different decision steps, DRL only needs to maximize the immediate reward \cite{ sunTwoStageVoltVar2021, caoDataDrivenMultiAgentDeep2021, nguyenThreeStageInverterBasedPeak2022}. 
We can simplify the state-action value function as
\begin{equation}\label{one_Q}
Q^{\pi}( s, a) = \underset{a \sim \pi}{\mathbb{E}}\left[  r \mid s, a \right].
\end{equation}
Then, the objective of RL is to find a policy $\pi$ to maximize the state-action value function:
\begin{equation}
    \pi^*=\arg \max_{a \sim \pi} Q^{\pi}(s,a).
\end{equation}

{\color{blue}
\subsubsection{Quantile Regression}

Quantile regression is a statistical technique that 
predicts multiple conditional quantiles of a response variable distribution rather than a single mean value of traditional mean regression \cite{koenkerQuantileRegression2005}. This approach provides a more comprehensive view of the uncertainty and variability in predictions.
The  $\tau^{th}$ quantile of the conditional distribution of the response variable $Y$ given the predictor variable $X$ can be described as 

\begin{equation}
f_{Y | X}(\tau) =  \inf \{y: F_{Y | X} \geq \tau\},
\end{equation}
where $\tau$ is the quantile, $F$ is the cumulative distribution function, and $f_{Y | X}(\tau)$ is the  ${\tau}^{th}$ conditional quantile value of $Y$ given $X$.

The $\tau^{th}$ quantile function $f$ can be obtained by solving the optimiation task:
\begin{equation}
\begin{split}
f_{Y|X}(\tau) = & \arg \min_{f} \mathbb{E}  \Big[(\tau-1)_{f_{Y|X}(\tau) > Y }\left( Y  - f_{Y|X}(\tau)\right)\\
&+\tau_{f_{Y|X}(\tau) < Y }\left( Y  - f_{Y|X}(\tau)\right)\Big]. 
\end{split}
\end{equation}

The quantile regression model predicts multiple quantile values to describe the distribution response variable $Y$ by setting multiple quantile $\tau$, e.g.,$ 0.1, 0.2, \dots, 0.9$. 
Note that  $\tau = 0.5$ indicates estimating the mean of the response variable, which is the same as the traditional least squares regression.
}

\section{Problem Formulation: Partially Observable Markov Decision Process}\label{POMDP}


The proposed robust DRL approach is model-free and learns to make decisions from limited measurements of ADNs. We consider the following measurement conditions:
1) Voltage magnitude measurements are deployed only on a few buses that need to be optimized;
2) Active and reactive power injection measurements are deployed on the root bus;
3) Branch current magnitude measurements are deployed only on a few distribution branches;
4) The active and reactive power load and the active power outputs of DGs are deployed only on a few buses.

IB-VVC of an ADN with a limited measurement deployment is formulated as {\color{blue} a partially observable Markov decision process (POMDP) $(\mathcal{S}, \mathcal{O}, \mathcal{A}, \mathcal{R}, \mathcal{\hat{R}})$,}  then is solved by the DRL approach.
$\mathcal{S}$ is the state space, $\mathcal{O}$ is observation space, $ \mathcal{A}$ is action space and $\mathcal{R}$ is the reward function.
Different from the traditional POMDP $(\mathcal{S}, \mathcal{O}, \mathcal{A}, \mathcal{R}, \mathcal{P})$ \cite{ghoshWhyGeneralizationRL2021, mengPartialObservabilityDRL2022} that assume the real reward $\mathcal{R}$ is known, there is an additional term: ``the surrogate reward $\mathcal{\hat{R}}$", to represent the real reward when the real reward is unknown.
{\color{blue} In addition, the proposed POMDP also omits the transition probability function $\mathcal{P}$ because the IB-VVC is a single period optimization task, and  DRL only needs to maximize the immediate reward and does not need to consider the future state \cite{ sunTwoStageVoltVar2021, caoDataDrivenMultiAgentDeep2021, nguyenThreeStageInverterBasedPeak2022}. This kind of formulation is also named the contextual bandit \cite{suttonReinforcementLearningSecond2018}.
}

At each time step, the state of the ADN environment is $s \in \mathcal{S}$, the DRL agent only observes a partial state $s_{o} \in \mathcal{O}$, and then outputs an action $a \in \mathcal{A}$ based on its policy $\pi: \mathcal{S} \rightarrow \mathcal{A}$.
After executing the action, the ADN environment transits to the next state $s^\prime$ with the corresponding real reward $r$, whereas the DRL agent
only observes a partially observable state $s^\prime_{o}$ and the surrogate reward $ \hat{r}$.
The DRL agent collects the data tuple {\color{blue} $(s_{o}, a_t, \hat{r}_t)$} into the data buffer for training the actor and critic networks.

 By referring to the IB-VVC formulation introduced in Section \ref{mb_IB_VVC}, the state, partially observable state, action, and reward of POMDP are defined as follows:

\subsection{State} The state reflects the working condition completely of the ADN, including the uncontrollable, controllable, and dependence variables. 
$s = (V, P_0, Q_0, I, P_{D}, Q_{D}, P_{G}, Q_G, Q_C)$.
In fact, the uncontrollable variables $(P_{D}, Q_{D}, P_{G})$ is enough for IB-VVC decision-making. We add the redundant measurements $ (V, P_0, Q_0, I, Q_G, Q_C)$ to reduce the learning difficulties of DRL  \cite{leiDataDrivenOptimalPower2021}.
Additionally, similar to the state estimation, the redundant information may represent the state robustly by eliminating the measurement error.
Note that the complete state can only be obtained for an ADN with enough measurements.

\subsection{Observation} 
For an ADN with limited measurement deployed, the observation is the partially observable state  $s_o = (V_o, P_0, Q_0, \\ I_o, P_{Do}, Q_{Do}, P_{Go}, Q_G, Q_C)$.
The subscript ``$_o$" indicates the observable measurements.
The vector of controllable variables $Q_G, Q_C$ is a linear mapping of the action of the DRL agent and does not need to be measured.

\subsection{Action} The action is related to the controllable variables, which is defined as $a = (a_{Q_G}, a_{Q_C})$.
For the convenience of constraining the controllable variables, the activation function of the output layer of the actor network is set as ``Tanh". Then the action $a$ is always in $(-1,1)$.
The controllable variables $Q_G$, $Q_C$ are obtained through linear mapping of $a$ from (-1, 1) to their output capability intervals.
We can represent it as $u = 0.5 ( \bar{u} - \underline{u}) a  + 0.5(\bar{u} + \underline{u})$, where $u$ is the concatenation of $Q_G, Q_C$, and $\underline{u}, \bar{u}$ are their upper and bottom bounds.
The action can also be considered as the normalized controllable variable $u$, which may reduce the learning difficulties of neural networks. 

\subsection{Reward}
The reward is to evaluate the optimization performance of the action, which {\color{blue}should be calculated based on the $P_0, P_G, P_D$ of the next state  $s^\prime$}.
It consists of two sub-terms: the reward of power loss and the reward of voltage violation.

The reward of power loss $r_{p}$ is defined as:
\begin{equation}\label{r_p}
r_p =  -(P_0 + \sum_{i \in \mathcal{N}_G}  P_{Gi} - \sum_{i \in \mathcal{N}_D}  P_{Di}).
\end{equation}

The reward of power loss can also be analytically calculated by
$r_p = - \sum_{i j \in \mathcal{E}}  r_{i j} \frac{P_{i j}^{\prime 2}+Q_{i j}^{2}}{V_{i}^ {2}}$, but it is seldom been used because it requires model parameters $r_{ij}$ that is difficult to be acquired.

The reward of voltage violation $r_{v}$ is defined as
\begin{equation}\label{r_v}
r_v=- \sum_{i \in \mathcal{N} } c_{vi}\left[\max \left(V_i-\bar{V}, 0\right)+\max \left(\underline{V}-V_i, 0\right)\right],
\end{equation}
where $c_{vi}$ is the weighting penalty coefficient of voltage violation. 
The negative sign is present because DRL aims to maximize the reward, which is converse with the optimization objectives of IB-VVC.




Equations \eqref{r_p} and \eqref{r_v} demonstrate that the reward of power loss is calculated based on the active power of the root bus $P_0$, the active powers of loads and generations $P_D$ and $P_G$. Note that although the latter two are not decision variables in the IB-VVC problem (1-5), they may vary in different samples collected at different times.
The reward of voltage violation is calculated from all bus voltages $V$. When all the active power injections and voltages are not measurable, the rewards for power loss and voltage violation become unknown, which limits the application of DRL. 

\subsection{Surrogate Reward}

For the unknown reward of power loss, we first propose the indirect reward and then propose the surrogate reward based on the indirect reward, which {\color{blue} are calculated based on the $P_0, V_o$ of the next state  $s_o^\prime$}.
The indirect reward is {\color{blue} the negative value of} active power of the root node,
\begin{equation}\label{p_r0}
\tilde{r}_p = - P_0.
\end{equation}
{\color{blue} 
We can also understand that the indirect reward ensures that power loss at unobservable nodes is also optimized from the perspective of model-based IB-VVC. For a given $P_D, Q_D, P_G$, when model-based IB-VVC decides a controllable reactive power $Q_G, Q_C$ resulting in a minimum power loss, then the active power output of root bus $P_0$ is also its minimum value because  $P_0  = - \sum_{i \in \mathcal{N}} P_{Gi}  + \sum_{i \in \mathcal{N}}  P_{Di} + P_{loss}$. This means that minimizing the power loss is equal to minimizing the active power injection of the root bus $P_0$.}
For the case of the partially observable state, even $P_{Gi}, P_{Di}$ may not be measurable; it can be inferred from the partially observable states \cite{mestavBayesianStateEstimation2019}.
For similar partially observable states, $P_{Gi}, P_{Di}$ tend to be similar. In those cases, if the DRL agent trials an action and obtains a smaller $P_0$ compared to other actions, it also indicates a smaller power loss.

Noting that $P_0$ includes the information of both $r_p$ and $\sum_{i \in \mathcal{N}} (P_{Di} - P_{Gi})$ and $r_p$ usually makes up only a small portion of $P_0$.
Approximating $P_0$ in DRL algorithms may be more challenging than approximating $r_p$.
It tends to result in a slightly larger approximation error of the critic network and degrade the VVC performance.
To alleviate the problem, the surrogate reward is designed as
\begin{equation}\label{p_r}
{\color{blue} \hat{r}_p} = \tilde{r}_p - \tilde{r}^{-}_p.
\end{equation}
{\color{blue} Note that $\tilde{r}_p = - P_0$ are calculated based on the element of the next partially observable state $s_o^\prime$, while $\tilde{r}^{-}_p = - P_0$ are calculated based on the element of the recent partially observable state $s_o$.}
For a given {\color{blue}partially observable} state $s_o$, $\tilde{r}^{-}_p$ is determined, if the DRL agent outputs an action $a$ that maximizes the surrogate reward $\hat{r}_{p}$, it also corresponds to the maximize value of $r_{p}$, and $\tilde{r}_{p}$.
In addition, the surrogate reward reflects the change of power loss directly, thus providing an apparent reward signal for DRL agents.



For the unknown reward of voltage violation, the corresponding surrogate reward can be calculated by those measurable bus voltages:
\begin{equation}\label{pr_v}
\hat{r}_v= -\sum_{i=\hat{\mathcal{N}}} c_{vi} \left[\max \left(V_i-\bar{V}, 0\right)+\max \left(\underline{V}-V_i, 0\right)\right].
\end{equation}
{\color{blue} The surrogate reward of voltage violation $\hat{r}_v$ are calculated based on the measurable voltage magnitude $V_o$ of the next partial observable state.}
Under the surrogate reward of voltage violation, the objective of DRL is to eliminate voltage violations of measurable buses directly.
Since the voltage in one bus has a mutual effect across other buses, voltages of the unmeasurable buses are optimized indirectly.


\begin{figure}[!ht]
\centering
\includegraphics[width=3in]{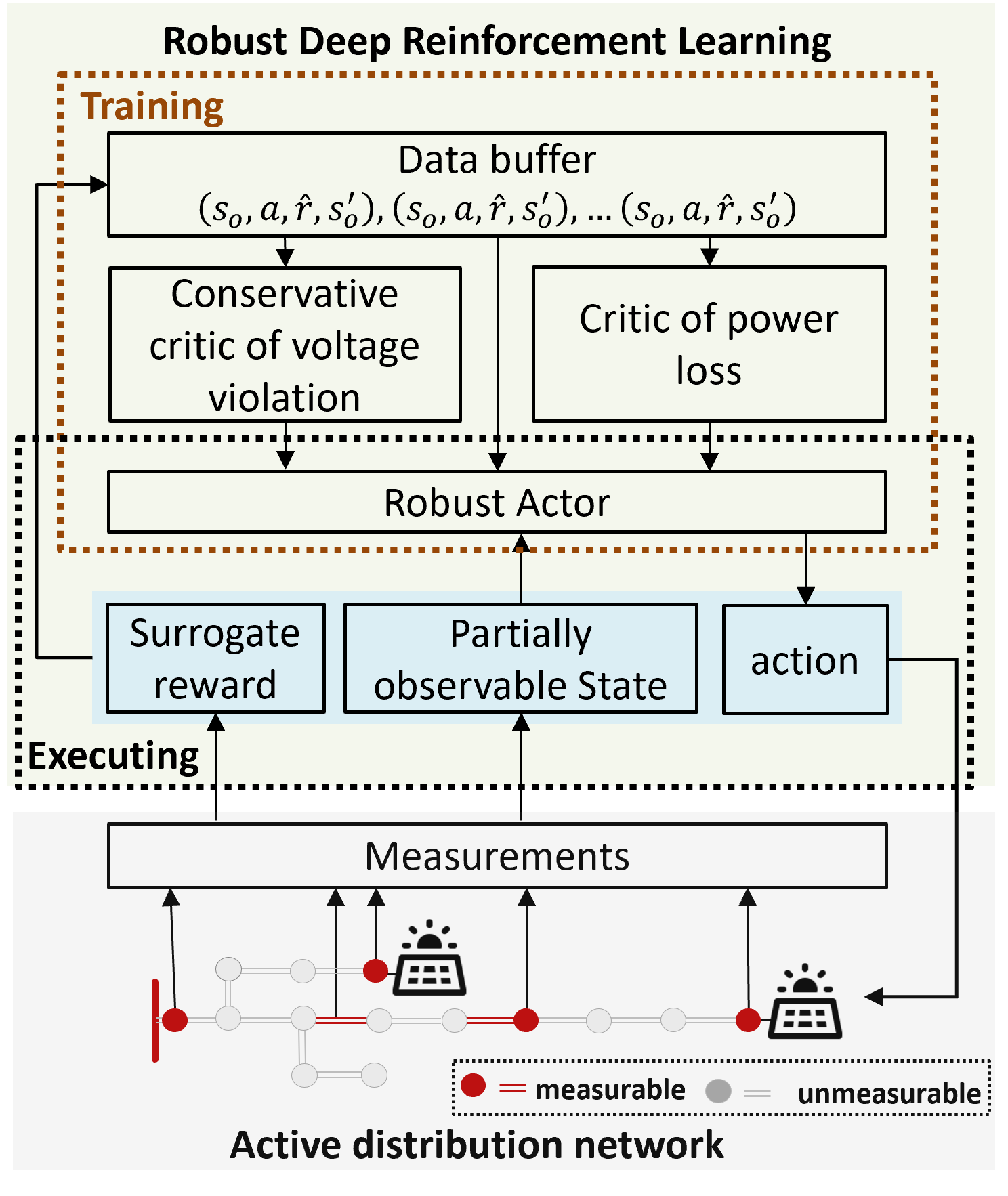}
\caption{The framework of robust deep reinforcement learning.
}\label{partial_DRL}
\end{figure}

\section{Robust Deep Reinforcement Learning Method}

The structure of the proposed robust DRL is shown in Fig. \ref{partial_DRL}.
In the training stage, a batch of historical data is sampled from the data buffer. The proposed robust DRL approach trains a conservative critic of voltage violation and a critic of power loss by using the sampling data.
Then, a robust actor is trained to output the action that maximizes the accumulative value of the two critics.
In the execution stage, the robust actor receives the partially observable state from the ADN environment and then outputs an action.
After executing the action in the ADN environment, the surrogate reward can be calculated based on available measurements.
The data tuple, consisting of the partially observable state, surrogate reward, and action, is then uploaded to the data buffer.

In this section, we first propose the conservative critic to address the problem of the partially observable state. Then, we design a practical robust DRL algorithm that integrates the state-of-the-art DRL algorithm with the proposed surrogate reward and conservative critic. Finally, we investigate the minimum measurement condition required for the proposed robust DRL approach.


\subsection{Conservative Critic}

In the case of partially observable states, there exists multiple state $s$ corresponding to the same partially observable state $s_o$.
We can use $s_o$ to predict $s$ but with a degree of uncertainty \cite{mestavBayesianStateEstimation2019}.




For the case with partially observable states, the critic network is 
\begin{equation}\label{Qp}
Q^{\pi}( s_o, a) = \underset{a \sim \pi}{\mathbb{E}} \left[  r \mid s, a \right],
\end{equation}
and the objective of DRL is finding {\color{blue} the optimal action $a^*$},
\begin{equation}
	a^* = \arg \max_a Q(s_o, a).
\end{equation}

Due to the uncertainties of predicting $s_o$ based on $s$, $\arg \max_a Q( s, a) \neq \arg \max_a Q(s_o, a)$ and it increases the risk of voltage violations.

There are two methods to address the problem. 
One method is adding more measurements into the partial observable state $s_o$ to improve the measurement condition, which has been studied in paper \cite{caoPhysicsInformedGraphicalRepresentationEnabled2024}.
When additional measurements are not available or insufficient, another method proposed in this paper is estimating the uncertainties of reward to the partially observable state, then constructing a conservative critic to estimate the conservative state-action value.
The robust policy can be trained with the guidance of the conservative critic network.

Inspired by paper \cite{dabneyDistributionalReinforcementLearning2018}, we use quantile regression technology to estimate the uncertainties of state-action value with respect to the partially observable state. 
Then state-action value \eqref{one_Q} is transformed to
\begin{equation}
Q_{r\mid(s,a)}(\tau) = \inf\{r: F_{ r\mid(s,a)} \geq \tau\},
\end{equation}
where $\tau$ is the quantile, $F$ is the cumulative distribution function, and $Q_{r\mid(s,a)}(\tau)$ is the  ${\tau}^{th}$ conditional quantile value of $r$ given $(s,a)$.
We set $\tau$ as a small value, like $0.2$ in this paper, to estimate the conservative sate-action value.
It can be estimated by solving the optimization problem:

\begin{equation}
\begin{split}
Q_{r\mid(s,a)}(\tau) = & \arg \min_{Q} \mathbb{E}  \Big[(\tau-1)_{Q(s_o, a) > r }\left( r  - Q(s_o, a)\right) \\
&+\tau_{Q(s_o, a) < r }\left( r  - Q(s_o, a)\right)\Big]
\end{split}
\end{equation}

Other uncertainty estimation technologies, like the Bayesian neural network, can solve the problem. Here, we select quantile regression because it has a simple structure and stable performance.

\subsection{A Practical Algorithm: Robust Soft Actor-Two-Critic}


The proposed robust soft actor-two-critic algorithm combines the soft actor-critic (SAC) \cite{haarnojaSoftActorCriticAlgorithms2019}, two-critic DRL framework \cite{liuTwoCriticDeepReinforcement2024}, conservative critic, and surrogate reward.
SAC is an off-policy actor-critic deep RL algorithm based on the maximum entropy reinforcement learning framework that provides a robust and sample-efficient learning performance.
It is a state-of-the-art DRL algorithm that has been widely utilized in VVC tasks \cite{wangSafeOffPolicyDeep2020,liuOnlineMultiAgentReinforcement2021, caoPhysicsInformedGraphicalRepresentationEnabled2024}.
The two-critic DRL framework is a specific scheme for IB-VVC, which uses two critic networks to approximate the two objectives of IB-VVC separately \cite{liuTwoCriticDeepReinforcement2024}. 
It reduces the learning difficulties of each critic and avoids the mutual effect of the optimization objectives in learning critic networks, thus accelerating the learning process and improving the VVC performance further.
In this paper, the two-critic DRL framework also provides a convenient way to design the conservative critic solely for voltage violations.

For the IB-VVC task, SAC only learns to maximize the entropy-regularized reward \cite{caoDataDrivenMultiAgentDeep2021, nguyenThreeStageInverterBasedPeak2022},
\begin{equation}
    \pi^{*}=\arg \max_{\pi} \underset{a \sim \pi}{\mathbb{E}} \left[ r+\alpha H\left(\pi\left(\cdot \mid s_{o}\right)\right)\right],
\end{equation}
where the entropy is $H\left(\pi\left(\cdot \mid s_{o}\right)\right)=\underset{a \sim \pi\left(\cdot \mid  s_{o}\right)}{\mathbb{E}}\left[-\log \pi\left(\cdot \mid  s_{o}\right)\right]$. 
Here, the partially observable state $s_o$ is utilized rather than $s$.

Integrating the two-critic DRL framework and the surrogate rewards, the critic of power loss $Q_p(s_o,a)$ and the critic of voltage violation $Q_v(s,a)$ are introduced: 
\begin{align}
Q_p^{\pi}( s_o, a) &= \underset{a \sim \pi}{\mathbb{E}}\left[  \hat{r}_p \mid s_o, a \right],\label{Qp} \\
Q_v^{\pi}( s_o, a) &= \underset{a \sim \pi}{\mathbb{E}}\left[ \hat{r}_v \mid s_o, a \right]
.\label{Qv}
\end{align}


In this paper, the objective of IB-VVC is to minimize the expected power loss and decrease the risk of voltage violation, so we only apply the conservative critic scheme to the critic of voltage violation.
Then, equation \eqref{Qv} is transformed to 
\begin{equation}\label{CQv}
Q^{\pi}_{\hat{r}_v \mid(s,a)}(\tau) = \inf\{r: F_{ \hat{r}_v \mid(s,a)} \geq \tau\},
\end{equation}
We set $\tau$ as a small value to have a conservative critic, like $0.2$ in this paper.

The two critics $Q_p\left(s_o, a \right)$ and $Q_v\left(s_o, a \right)$ are parameterized by two neural networks $Q_{\phi_p}$ and $Q_{\phi_v}$ with learnable parameters of $\phi_p$ and  $\phi_v$.
{\color{blue} The critic of power loss} $Q_{\phi_p}(s_o, a)$ in \eqref{Qp} is
learned by minimizing the MSE losses
\begin{equation}\label{p_QMSE}
L_{Q_p}(\phi_p) = \frac{1}{|B|} \sum_{\left(s_o, a, \hat{r}_p \right) \in B}\left(Q_{\phi_p}(s_o, a)-r_p\right)^{2}.
\end{equation}
{\color{blue} The critic of voltage violation} $Q_{\phi_v}(s_o, a)$ is learned by minimizing the quantile regression loss
\begin{equation}\label{p_VMSE}
\begin{split}
L_{Q_{v}}(\phi_v) =&  \Big[(\tau-1)_{Q_{\phi_v}(s_o, a)>\hat{r}_v}\left( \hat{r}_v - Q_{\phi_v}(s_o, a)\right) \\
&+\tau_{Q_{\phi_v}(s_o, a) < \hat{r}_v}\left( \hat{r}_v - Q_{\phi_v}(s_o, a)\right)\Big].
\end{split}
\end{equation}

The actor is a stochastic policy, which is parameterized as
\begin{equation}\label{stochastic_policy}
    \pi_{\theta}(\cdot \mid s_o)=\tanh \left(\mu_{\theta}(s_o)+\sigma_{\theta}(s_o) \odot \xi\right), \quad \xi \sim \mathcal{N}(0, I),
\end{equation}
where $\mu_{\theta}$, $\sigma_{\theta}$ are neural networks to approximate mean and standard deviations, {\color{blue}and $\xi$ is a random variable generated from a standard Gaussian distribution $\mathcal{N}(0, I)$}.
The actor network $\pi_{\theta}(\cdot \mid s)$ learn to maximize the entropy regularized critic value, and the loss function is 
\begin{equation}\label{twosac}
\begin{split}
L_{\pi}({\theta}) =   & \frac{1}{|B|} \sum_{s_o \in B, a \sim  \pi_{\theta}(\cdot \mid s_o) } \big( Q_{p_{\phi}}\left(s_o, a \right)  +  Q_{v_{\phi}}\left(s_o, a \right)   \\
& -  \alpha \log \pi_{\theta}\left(\cdot \mid s_o \right) \big),
\end{split}
\end{equation}
where $\log \pi_{\theta}\left(\cdot \mid s_o \right)$ is the entropy of the policy, and  $\alpha$ is the temperature parameter. $\alpha$ can be a constant or adjusted by minimizing the loss
\begin{equation}
    L(\alpha) =  \frac{1}{|B|} \sum_{s_o \in B  }  \big(-\alpha \log \pi_{\theta} (\cdot |s_o) - \alpha \mathcal{H} \big), 
\end{equation}
where $\mathcal{H}$ is the target entropy.

\begin{algorithm}[!t]
  \caption{Robust Soft Actor-Two-Critic} \label{RSATC}
  \begin{algorithmic}[1]
  \Require
  Initialize the critic network parameters  $\phi_p$, $\phi_v$, the actor network parameters $\theta$, empty replay buffer $\mathcal{D}$;
\For {each environment step}
\State Execute $a \sim \pi(\cdot \mid s_o )$ in the ADN environment;
\State Observe a reward $r_p, r_v$ and the next state $s^\prime$;
\State Store $(s,a,r_p, r_v, s^\prime)$ in replay buffer;
\For {$j$ in range (how many updates)}
 \State Randomly sample a batch data $B$ from $\mathcal{D}$;
\State Update $Q_{\phi_p}$ and $Q_{\phi_v}$ according to
$$
\phi_p \leftarrow \phi_p - \lambda_Q \nabla_{\phi_p} L_{Q_p}(\phi_p)
$$
$$
\phi_v \leftarrow \phi_v - \lambda_Q \nabla_{\phi_v} L_{Q_v}(\phi_v)
$$
\State Update $\pi_{\theta}$ according to
$$
\theta \leftarrow \theta + \lambda_{\pi} \nabla_{\phi} L_{\pi}(\theta)
$$
\EndFor
\EndFor
\end{algorithmic}
\end{algorithm}


The proposed approach optimizes the power loss of the whole network and the voltage profile of buses with measurable voltages while indirectly improving the voltage profile of other buses.
The other buses that are closer to the measurable bus would have better voltage optimization performance.
The configuration of the measurable bus is decided by the distribution system operator, and voltage sensors would be installed on the pilot nodes or nodes with significant users.
The implementation details of the proposed robust soft actor-two-critic algorithm are provided in Algorithm \ref{RSATC}.

{\color{blue}
Deciding the weighting penalty coefficient of voltage violation $c_v$ is crucial for the proposed DRL approach. Since DRL is a model-free approach, we need to tune the $c_v$ to achieve an acceptable performance.
A large $c_{v}$ would decrease the voltage violation rate while increasing the power loss, whereas a small $c_v$ can not eliminate voltage violations successfully. 

In our simulation, empirically, we think the value of $c_v$ is suitable when we observe the following two phenomenons:
\begin{itemize}
\item[1.] During the initial stage, the rate of voltage violations decreases rapidly and approaches zero. This suggests that $c_v$ is sufficiently large to optimize the voltage violations.
If the rate of voltage violations cannot approach zero, we need to increase the value of $c_v$.
If $c_v$ is very large, e.g. $1000$, and the rate of voltage violations still cannot approach zero, we may need to check whether the VVC problem has feasible solutions.
\item[2.] After enough time to learn, about $5-10\%$ episodes exhibit zero voltage violation, and others have a low voltage violation rate. 
The low voltage violation rate is caused by the optimization objectives of power loss and voltage violation conflicting in some circumstances. 
In the training trajectory, DRL algorithms must trail both sides of the voltage boundary many times to approach the optimal solution.
If there are larger $5-10\%$ episodes exhibiting zero voltage violation, we may need to decrease $c_v$ slightly, and vice versa.
\end{itemize}
}

\begin{remark}
{\color{blue} We wish to clarify that the two-critic scheme proposed in paper \cite{liuTwoCriticDeepReinforcement2024} does not contravene the original intent of having two critics in SAC, but is also compatible with SAC \cite{haarnojaSoftActorCriticAlgorithms2019}. 
Different from the two-critic scheme \cite{liuTwoCriticDeepReinforcement2024}, the original intent of two critics in SAC is to use the smaller of the two critic values to form the targets in the Bellman error loss functions to address the overestimation of critic value, also known as ``clip double Q learning".
We can apply the two-critic scheme \cite{liuTwoCriticDeepReinforcement2024} to the standard SAC with clip double Q learning. 
In this paper, we did not use the clip double Q learning because IB-VVC is a single-period optimization, and we modified the SAC to maximize the immediate reward rather than the discounted accumulated reward.
}
\end{remark}

\subsection{The minimum measurement condition}


The proposed robust DRL approach is applicable to different measurement conditions.
The minimal measurement condition is the measurement of the active power injection of the root bus $P_0$ and a few bus voltage magnitudes $V_o$ that need to be optimized. Those measurements can be used as the observation for the proposed DRL algorithm and are also enough to calculate the surrogate reward.
Under minimal conditions, DRL algorithms still can work well, which will be verified in the following simulation.

\section{Simulation}
We conducted numerical simulations on 33, 69, and 118-bus distribution networks to demonstrate the advantages of the proposed robust DRL approach.
{\color{blue} Those distribution network data were downloaded from Matpower \cite{zimmermanMATPOWERSteadyStateOperations2011} and then converted to Pandapower \cite{thurnerPandapowerOpenSourcePython2018}.}
The configurations of DGs and SVCs were shown in Table \ref{simulation_setting}.
{\color{blue} There were 2, 4, and 8 DGs, and 1, 1, and 2 SVCs in 33, 69, and 118-bus, respectively.}
The reactive power capacities of all DGs and SVCs were set as 3 and 2 MVar.
All load power varies between $50\%$ and $150\%$ of their default values following uniform distributions.
The active generations were a uniform distribution in the interval $[0.5,1.5]$ MW in the 33 and 69 bus systems and $[1,2]$ MW in the 118 bus system. 
The lower and upper bounds of the voltage
magnitudes are set as 0.95 and 1.05 p.u., respectively. 

\begin{table}[!t]
\renewcommand{\arraystretch}{1.3}
\caption{The configurations of DGs and SVCs in distribution networks}
\label{simulation_setting}
\centering
\begin{tabular}{ c c c}
\hline
 \text {Network } & \text { Devices } & \text { buses } \\
\hline
\multirow{2}{*}{\text{33-bus}} & \text {DG} &  15, 24  \\
& \text {SVC} & \text { 32 } \\
\hline 
\multirow{2}{*}{\text{69-bus}} & \text {DG} &   5,22,44,63  \\
& \text {SVC} & \text { 52 } \\
\hline 
\multirow{2}{*}{\text{118-bus}} & \text {DG} &  33, 44, 50, 53, 76, 97, 106, 111 \\
& \text {SVC} & \text 69, 84 \\

\hline 
\end{tabular}
\end{table}

\begin{table}[!t]
\centering
\begin{threeparttable}
\renewcommand{\arraystretch}{1.3}
\caption{The configurations of measurements}
\label{observation_case}
\centering
\begin{tabular}{cccccc}
\hline
 & &  & Partial state &  & Complete state\\
\hline
\multicolumn{1}{c|}{\multirow{4}{*}{$O_3$}} & \multicolumn{1}{c|}{\multirow{3}{*}{$O_2$}} & $O_1$  & $P_0, V_{o}$ &  & $P_0,V$  \\
\cline {4-6}
\multicolumn{1}{c|}{}  & \multicolumn{1}{c|}{}  &  & 
 $I_o$ &  &  $I$ \\
  \cline {4-6}
\multicolumn{1}{c|}{}   & & & \multirow{2}{*}{\makecell[c]{$Q_0, P_{Do},$ \\ $ Q_{Do}, P_{Go}$}}   & &  \multirow{2}{*}{\makecell[c]{$Q_0, P_{D},$ \\ $ Q_{D}, P_{G}$}} \\
\multicolumn{1}{c|}{} &   &  & &  \\

\hline 
\end{tabular}

\end{threeparttable}
\end{table}

As shown in Table \ref{observation_case}, three measurement conditions were tested to verify the effectiveness of the proposed robust DRL.
1) \textit{O1:} O1 is the minimal measurement condition that only the active power of the root bus and the voltage magnitudes of a few buses were measurable. Here, we set the voltages in buses with DGs and SVCs to be measurable.
2) \textit{O2:} Measurements in O1 plus some branch current magnitude measurements.
There were 5, 8, and 16 branch current magnitude measurements in 33, 69, and 118 bus distribution networks, respectively.
3) \textit{O3:} Measurements in O2 plus the active and reactive power of loads in the buses with DGs and SVCs and the active power generations of DGs.
The detailed configuration can be seen in our open source code \cite{qiongIkelqRDRL_PO2024}.
Three corresponding complete states are also designed. In three complete states, all corresponding voltages, active and reactive power of loads, the active power generations of DGs, and branch currents are measurable.

\begin{table}[!t]
\renewcommand{\arraystretch}{1.3}
\caption{Parameter setting for the reinforcement learning algorithms}
\label{DRL_paramter}
\centering
\begin{tabular}{ c c}
\hline
 \text { Parameter } & \text { Value } \\
\hline 
 \text { Optimizer } & \text { Adam } \\
\text { Activation function } & \text { ReLU } \\
 \text { Number of hidden layers } & 2  \\
 Actor hidden layer neurons of actor and critic & \{512, 512 \} \\
 \text { Batch size } &  128 \\
 \text { Replay buffer size } & $ 3 \times 10^{4}$ \\
 \text {Critic learning rate} &  $ 3 \times 10^{-4}$\\
 \text {Actor learning rate   } &  $ 1 \times 10^{-4}$\\
 Voltage violation penalty $c_{vi}$ & 50\\
 Initial random step & 1000\\
 Iterations per time step & 4\\
Entropy target & $-\dim(\mathcal{A})$\\
Temperature learning rate & $ 3 \times 10^{-4}$ \\
Quantile & $0.2$ \\
\hline 
\end{tabular}
\end{table}

\subsection{Effectiveness of the Proposed Robust DRL} \label{overall_performance}

This section conducts the following simulation to demonstrate the effectiveness of the proposed robust DRL algorithm for ADNs with limited measurement deployments:
\begin{itemize}
    \item[1.] The standard DRL approach is under the ideal measurement conditions. All measurements including $V, P_0, Q_0, I, P_{D}, Q_{D}, P_{G}, Q_G, Q_C$ are available. This means that the approach utilizes the complete state and complete reward (CC).
    \item[2.] The proposed robust DRL approach is under three limited measurement conditions. This means that the approach utilizes the partial observable state and surrogate reward (PS). The three simulations are donated as PS-1, PS-2, and PS-3, respectively.
    \item[3.]
        {\color{blue} Model-based optimization was formulated based on the AC power flow model and solved by recalling PandaPower \cite{thurnerPandapowerOpenSourcePython2018} with the interior point solver.
        To be consistent with the DRL approaches, in model-based optimization, the voltage constraints were only added to the buses with voltage measurements.}
        The result can be seen as optimal, which is a baseline for the performance of DRL algorithms. Note that the approach only works on the ideal measurements and ideal power flow model condition.
    \item[4.] The without-control strategy (W) sets the reactive power generation of DGs and SVC at zero.
\end{itemize}
All of the simulation experiments are tested on 33, 69, and 118 bus distribution networks to show the scalability of the proposed method.
Python was used to implement the algorithms, PyTorch was used for the DRL algorithms, and Pandapower \cite{thurnerPandapowerOpenSourcePython2018} was used to calculate the balanced power flow for simulating the ADN environments. 
{\color{blue} We trained the DRL agent using 100000 data. The parameter setting for the DRL algorithms is provided in Table \ref{DRL_paramter}.
In the training process, we tested the DRL algorithms in the same environment at each step. }

\begin{figure*}[ht]
\centering
\subfloat[33-bus]{
    \includegraphics[width=2.3in]{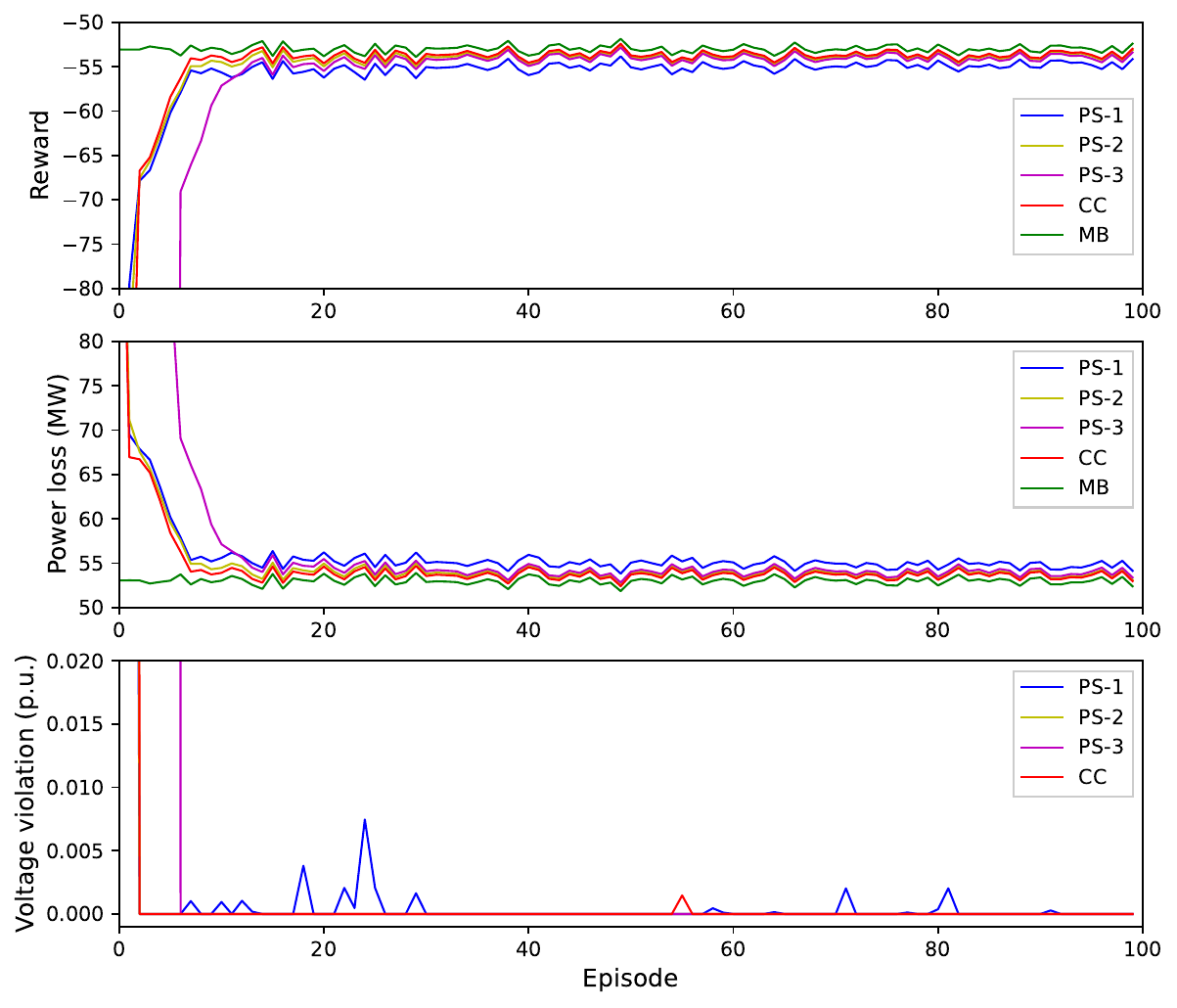}
    }
\subfloat[69-bus]{
\includegraphics[width=2.3in]{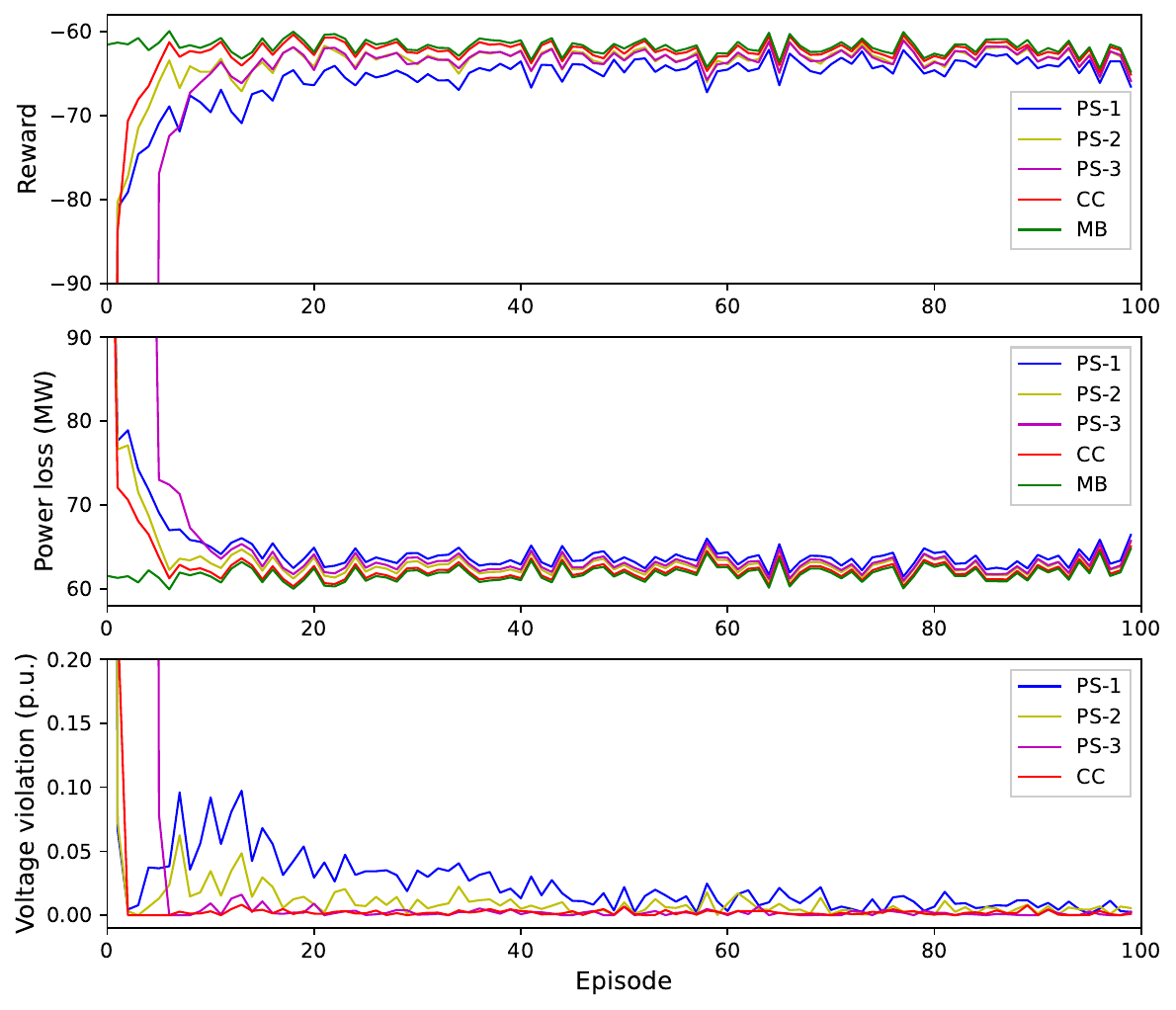}
    }
\subfloat[118-bus]{
\includegraphics[width=2.3in]{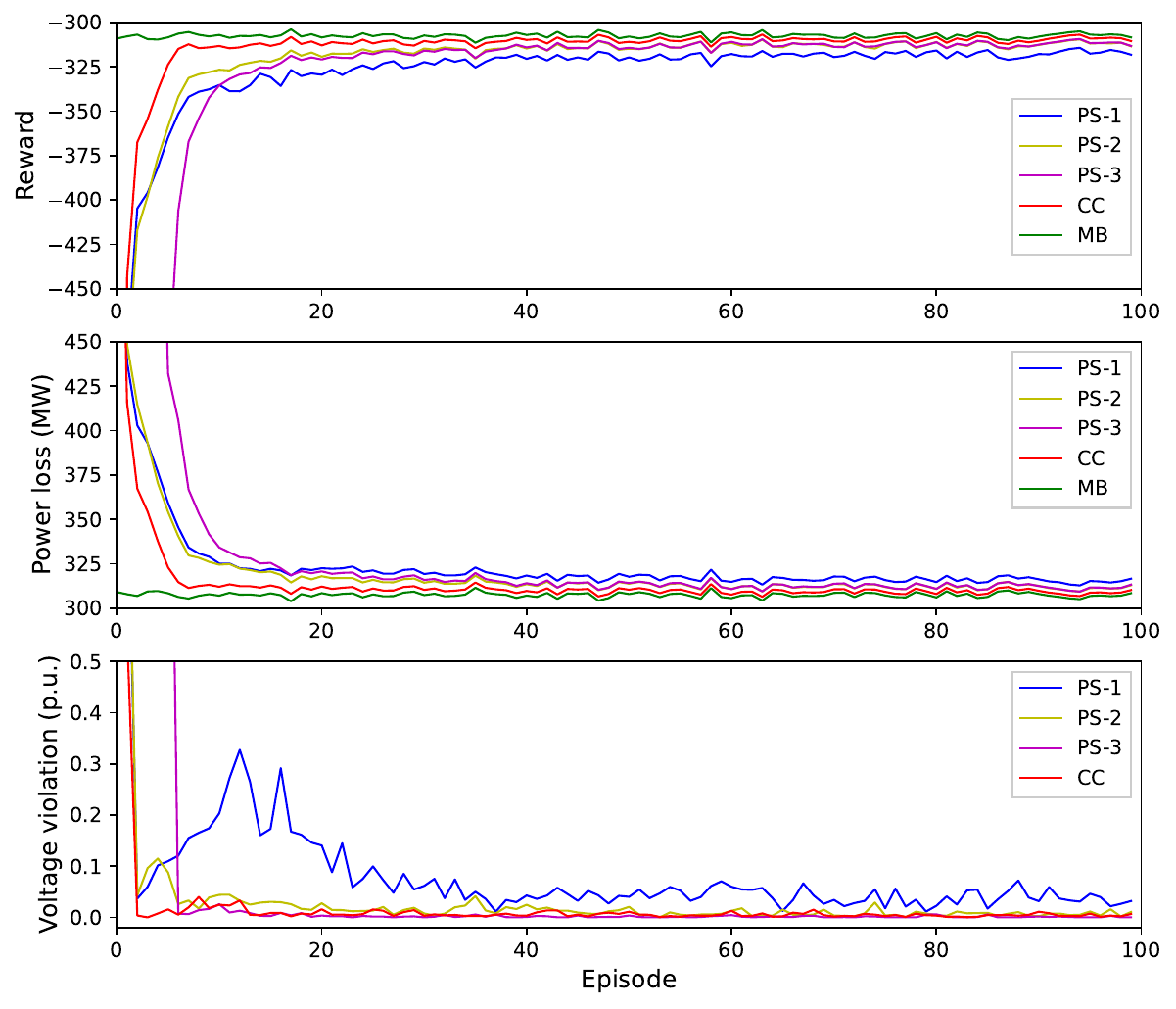}
    }
\caption{The testing results of the proposed robust DRL with partially observable state and surrogate reward (PS) under three measurement conditions, the standard DRL with complete state and complete reward (CC), and the model-based optimization (MB).
The minimal measurement condition, additional branch currents, and additional power measurements of the few buses and a few branch currents were symbolized as O1, O2, and O3. The results were plotted from the 11th episode.
The value of the reward, power loss, and voltage violation is the accumulated value of one episode.
}\label{DRL_result}
\end{figure*}


Fig. \ref{DRL_result} presents the simulation results of the standard DRL algorithm with complete state and complete reward (CC), the proposed robust DRL algorithm with three types of partial observable state and surrogate reward (PS-1, PS-2, PS-3), and the model-based optimization approach with sufficient measurements and an accurate model.
In the simulation results, we calculate the reward for power loss using equation \eqref{r_p}. We calculate the surrogate reward for power loss using equation \eqref{p_r}.
We calculate the reward for voltage violation for all experiments of DRL approaches using equation \eqref{pr_v}.
During the experiments, the reward, power loss, and voltage violation were the accumulated values in each episode.
Each episode consisted of 1000 steps.
{\color{blue}The model-based approach also followed the rules of DRL approaches to calculate the reward, power loss, and voltage violation.}
The voltage violations of the model-based optimization approach were always zeros, which were omitted in Fig. \ref{DRL_result}.

We made three observations from Fig. \ref{DRL_result}:

{\color{blue} First, the model-based approach under the ideal measurements and ideal power flow model achieved the best VVC performance in terms of reward, power loss, and voltage violation. 
The reward value for the model-based (MB) algorithm exhibits fluctuations because the load and generation profiles are randomly generated and are different in each episode. The power loss would change with the load and generation profiles.
}

First, the proposed robust DRL under three partial observable conditions (the blue, yellow, and magenta lines) converged close to the standard DRL with ideal measurement conditions (the red line) regarding reward, power loss, and voltage violation, even on the minimal measurement condition, where only 4, 6, and 11 measurements were in 33, 69, and 118 bus distribution networks.
The result was consistent among the three distribution networks and the three limited measurement conditions.
It indicates that DRL can effectively make decisions by utilizing the partially observable state and surrogate reward, even when only the active power of the root bus and less than $10\%$ of bus voltages are measurable.

Second, the VVC performance of the robust DRL algorithms improves as the measurement condition increases.
By adding a few branch current magnitude measurements based on the minimal measurement condition, the VVC performance and learning efficiency improved, which can be seen by comparing the results of the proposed robust DRL under O1 (the blue line) with the corresponding under O2 (the yellow line).
Additionally, measuring reactive and reactive power injections of a few buses further enhances VVC performance and learning efficiency, which can be seen by comparing the results of the proposed robust DRL under O2 (the yellow line) with the corresponding under O3 (the magenta line).
These results were apparent across three distribution networks.
We can also find that as the number of measurements increases from O1 to O2 to O3, the performance improves, but the improvement rate may decrease.

Third, DRLs with bad observation conditions tend to have unstable learning processes.
In the minimal measurement conditions (the blue line), the trajectories of voltage violation in 69 and 118 bus systems fluctuate significantly in the initial learning process.
The behavior would be alleviated by generating more data.

\begin{figure}[ht]
\centering
\subfloat[33-bus]{\begin{minipage}[t]{0.36\linewidth}\label{compared_without_control33}
    \includegraphics[height=3.05in]{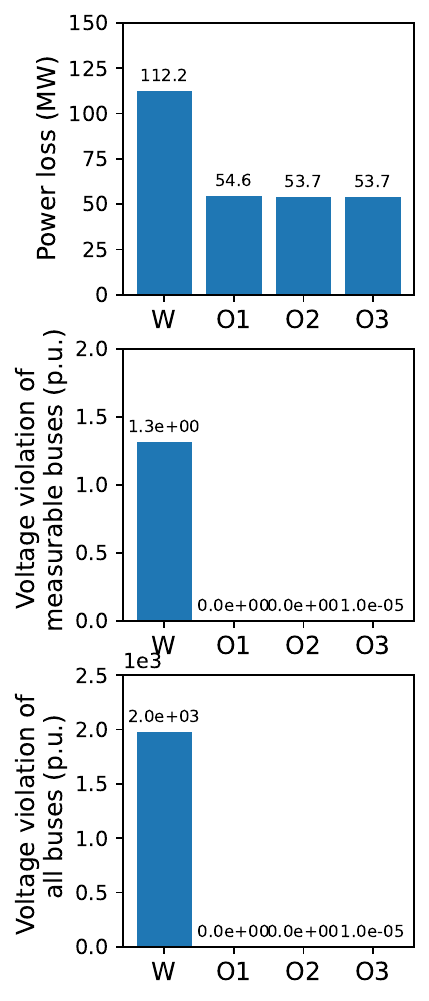}
    \end{minipage}
    }
\subfloat[69-bus]{
\begin{minipage}[t]{0.31\linewidth}\label{compared_without_control69}
\includegraphics[height=3.05in]{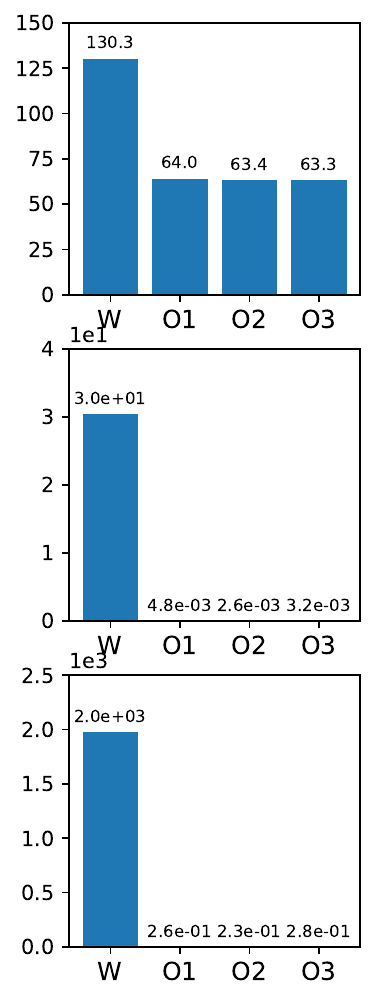}
\end{minipage}
    }
\subfloat[118-bus]{
\begin{minipage}[t]{0.31\linewidth}\label{compared_without_control118}
\includegraphics[height=3.05in]{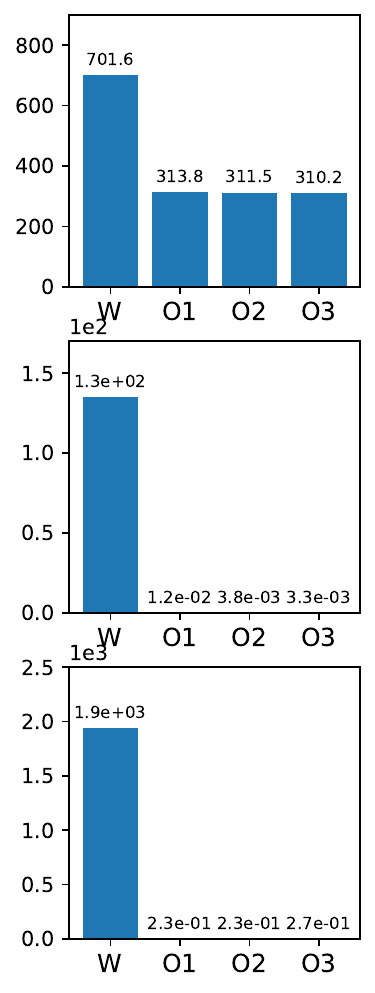}
\end{minipage}
    }
\caption{The VVC performance of without control and the robust DRL robust approach. The figure showed the mean value of the final 20 episodes.``W", ``O1", ``O2", ``O3"  represent the experiment without control and robust DRL on three limited measurement conditions (``O1", ``O2", ``O3"). }\label{compared_without_control}
\end{figure}


\begin{figure*}[ht]
\centering
\subfloat[33-bus]{\begin{minipage}[t]{0.32\linewidth}\label{error_result_qr}
    \includegraphics[width=2.3in]{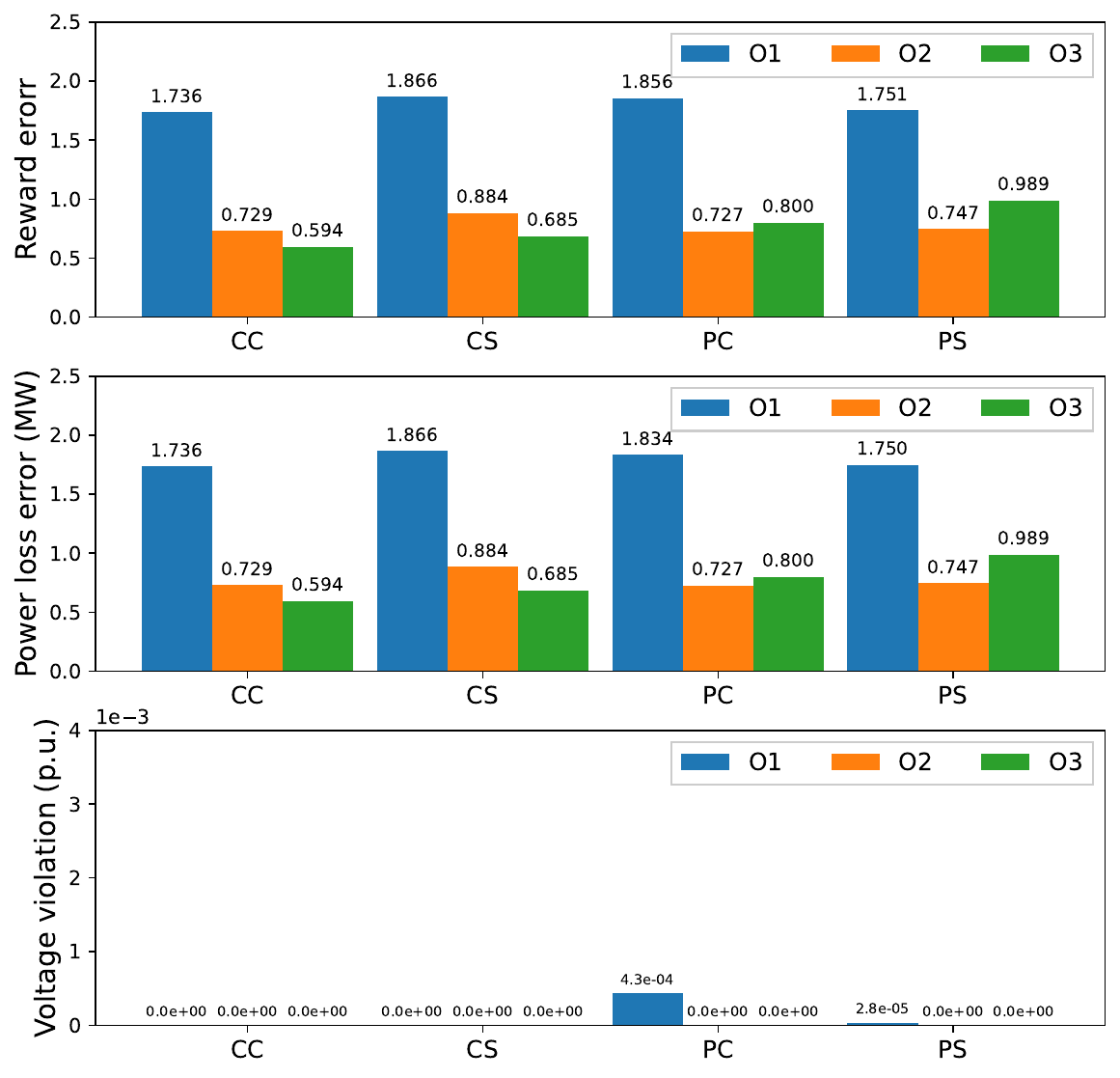}
    \end{minipage}
    }
\subfloat[69-bus]{
\begin{minipage}[t]{0.32\linewidth}\label{DRL_result_69}
\includegraphics[width=2.3in]{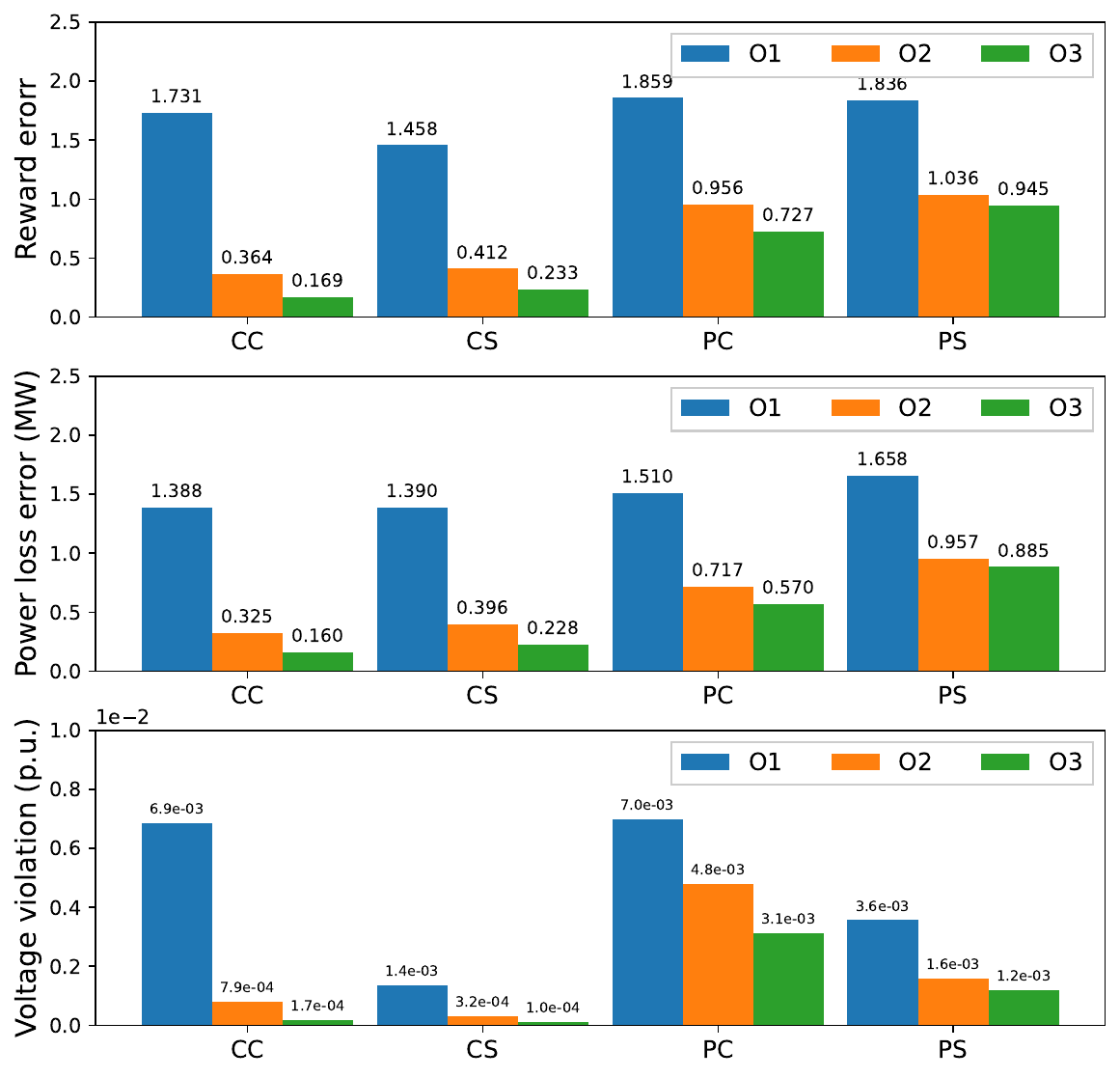}
\end{minipage}
    }
\subfloat[118-bus]{
\begin{minipage}[t]{0.315\linewidth}\label{DRL_result_118}
\includegraphics[width=2.3in]{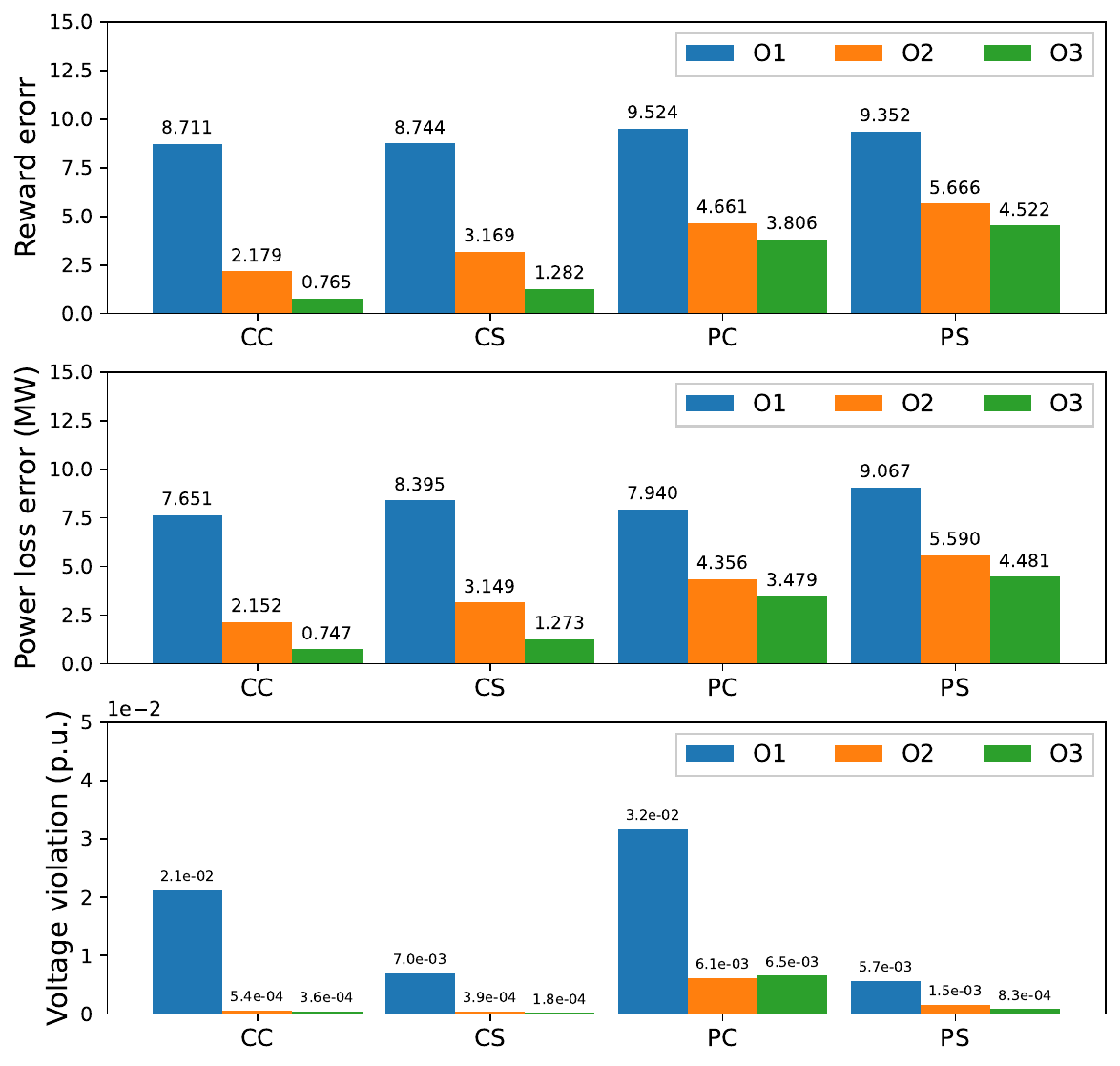}
\end{minipage}
    }
\caption{The optimization error in the final 20 episodes. Optimization error = optimization result of MBO - optimization result of the mentioned method. We test the proposed DRL approach under complete state and complete reward (CC), complete state and surrogate reward (CS), partially observable state and complete reward (PC), and partially observable state and surrogate reward (PS) with three measurement conditions (``O1", ``O2", ``O3") and three distribution networks.}\label{DRL_result_qr}
\end{figure*}

\begin{figure*}[ht]
\centering
\subfloat[33-bus]{\begin{minipage}[c]{0.325\linewidth}\label{error_result_33_O1}
    \centering
    \includegraphics[width=2in]{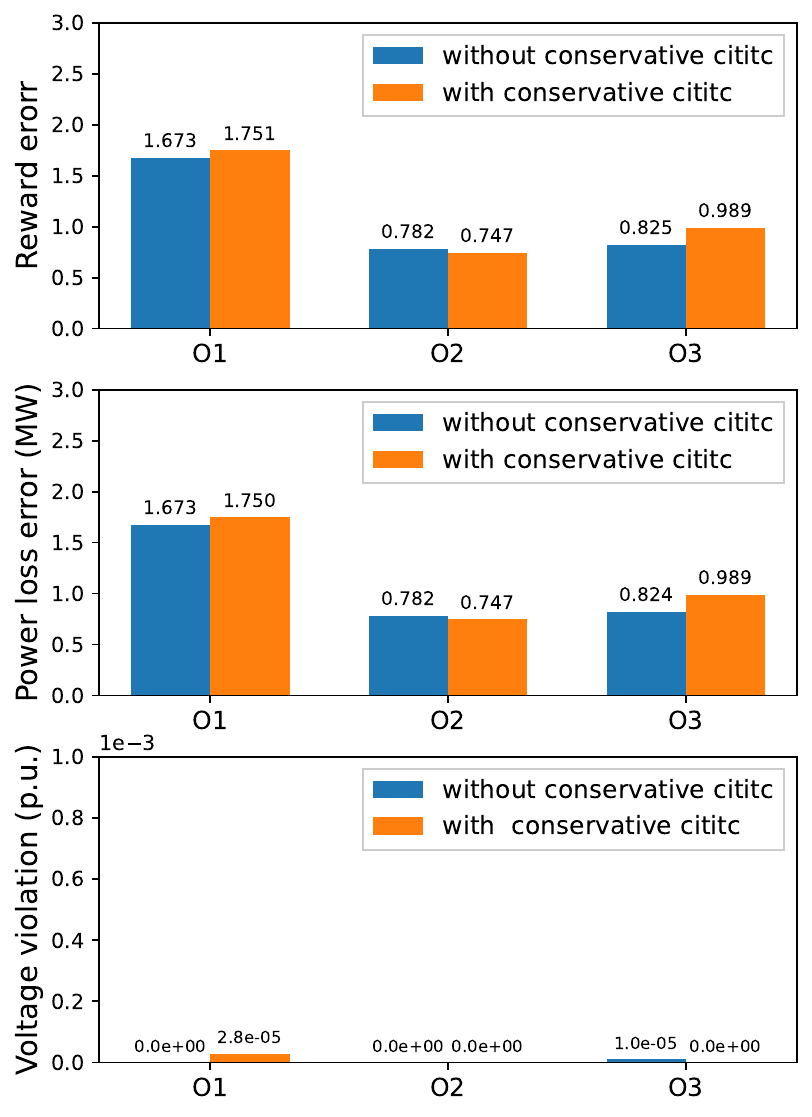}
    \end{minipage}
    }
\subfloat[69-bus]{
\begin{minipage}[c]{0.325\linewidth}\label{error_result_69_O1}
\centering
\includegraphics[width=2in]{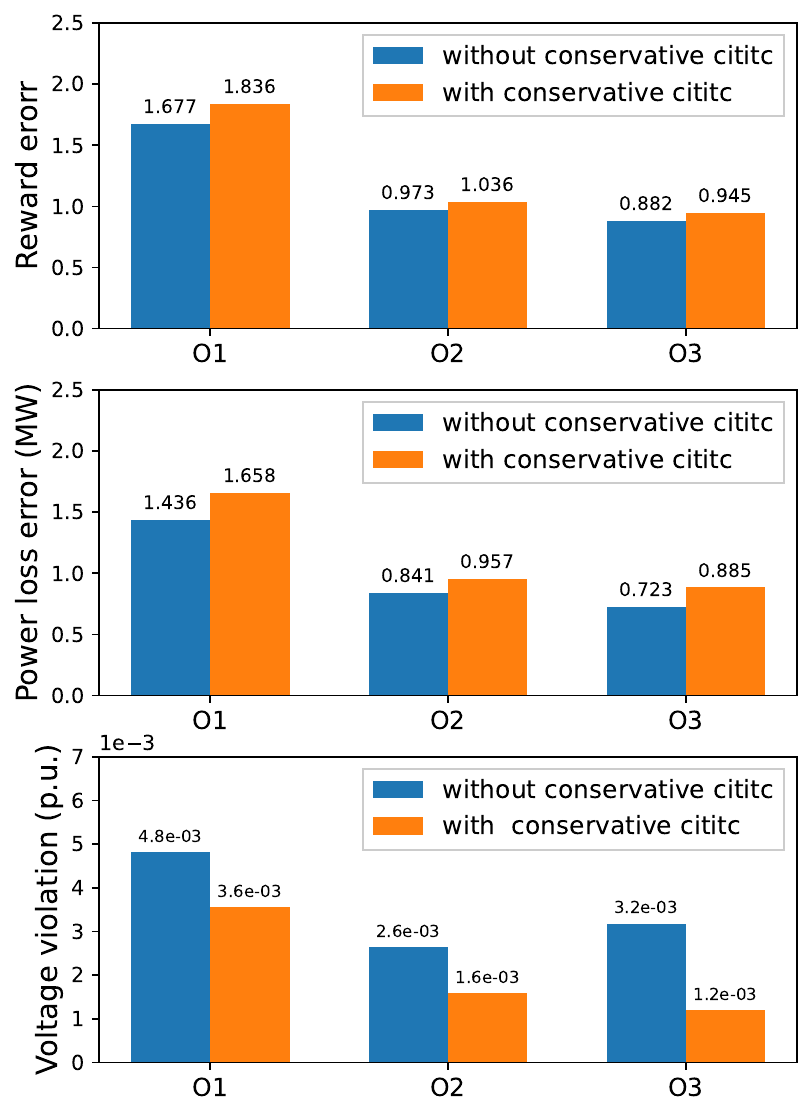}
\end{minipage}
    }
\subfloat[118-bus]{
\begin{minipage}[c]{0.325\linewidth}\label{error_result_118_O1}
\centering
\includegraphics[width=2in]{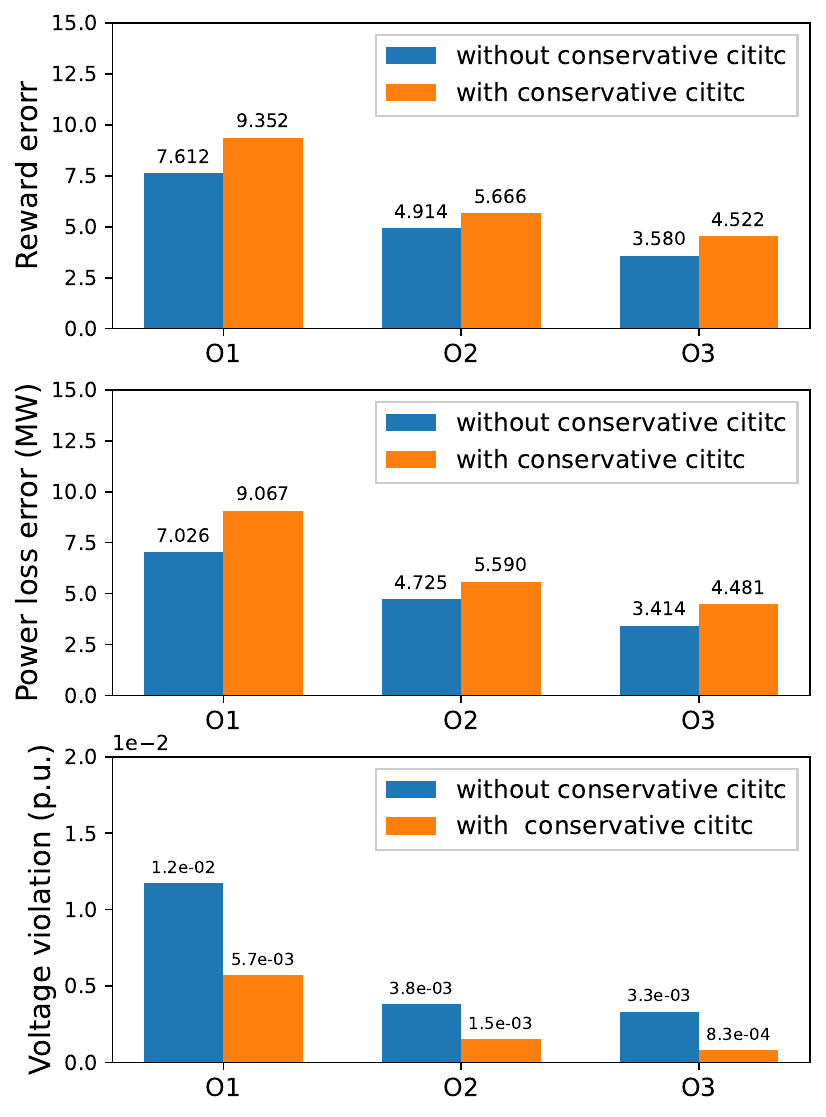}
\end{minipage}
    }
\caption{The optimization error DRL with and without the conservative critic in the final 20 episodes.}\label{error_result_C}
\end{figure*}

Fig. \ref{compared_without_control} showed the VVC performance of the experiments without control and the proposed robust DRL approach. We made two observations:

First, the proposed DRL approaches under three limited measurement conditions (legend “O1”, “O2”, and “O3”) achieved significantly better performance compared with the experiments without control (legend "W") in terms of power loss, voltage violation of measurable buses, and of all buses. It demonstrated that the proposed robust DRL for ADNs with limited measurement deployments can optimize the power loss of the whole network and the voltage profile of buses with measurable voltages while indirectly improving the voltage profile of other buses.

Second, the proposed DRL approach was robust to three limited measurement conditions. As the measurement condition decreased from "O3" to "O2" to "O1", the VVC performance of the proposed DRL only decreased slightly in terms of power loss, voltage violation of measurable buses, and all buses. 


\subsection{The Effect of Partially Observable State and  Surrogate Reward}

To demonstrate the impact of partial observable reward and surrogate reward, we tested the three measurement conditions mentioned above with both surrogate reward and complete reward.
Therefore, we designed the four types of simulation settings:
1) complete state and complete reward (CC),
2) complete state and surrogate reward (CS),
3) partially observable state and complete reward (PC),
and 4) partially observable state and surrogate reward (PS).
In those DRL approaches, we calculate the reward for voltage violation using equation \eqref{pr_v}.




Comparing simulation results between CC and PC and between CS and PS in Fig \ref{DRL_result_qr} showed the effect of the partially observable state.
It was apparent that the partially observable state degraded the VVC performance and slightly increased the power loss and voltage violation rate.
We also observed that increasing measurement conditions from O1 to O3 of PC or PS improves VVC performance in power loss and voltage violation.
Starting from the minimal measurement condition (O1), VVC performance increased when additional branch current magnitudes were added into the state.
Further improvements were achieved by adding additional power injection of a few buses into the state.
Generally, the improvement rate from O1 to O2 was larger than that of O2 to O3.
There were only two exceptions when comparing the power loss of PC-O2 VS PC-O3, and PS-O2 VS PS-O3 in the 33 bus distribution system, and one exception when comparing the voltage violation of PC-O2 VS PC-O3 in the 118 bus distribution system.

For complete states, adding a few branch currents and additional power injection of a few buses also enhances VVC performance.
The results were consistent among all experiments when comparing experiments O1, O2, and O3 between CC and CS.
It verified that the redundant information reduces DRL's learning difficulties.  
Compared with paper \cite{leiDataDrivenOptimalPower2021} that only verified the rule that uses bus voltages as redundant information in supervised learning paradigms, this paper also verified the effectiveness of using branch currents as redundant information in DRL paradigms.



Comparing the simulation results of CC vs. CS and PC vs. PS in Figure \ref{DRL_result_qr} showed the effect of the surrogate reward of power loss.
We found that the power loss error in CC and PC was slightly higher than the corresponding CS and PS. 
This verified that the proposed approach can optimize the power loss of the whole network.
Reasonably, due to limited measurement deployments, the VVC performance experienced a slight degradation.

The simulation results in Figure \ref{compared_without_control} showed the effect of surrogate reward. 
We observed that the voltage violation of measurable buses was approaching zero. Although the voltage violation of all buses was slightly higher than that of measurable buses, it was still significantly lower than the voltage violation of the experiments without control.
It verified that DRL, with the surrogate reward of voltage violation, optimized the voltage violation of measurable buses directly and of other buses indirectly.


\subsection{The Effect of Conservative Critic}

To demonstrate the effectiveness of the conservative critic, we tested the proposed robust DRL and the standard DRL under three limited measurement deployment conditions. 
The simulation results were shown in Figure \ref{error_result_C}.
We observed that the conservative critic reduced the voltage violation value considerably but increased the power loss slightly.
The results were consistent when comparing the experiments of DRL with and without the conservative critic on the three limited measurable conditions.


\section{Conclusion}
This paper proposes a robust DRL approach for IB-VVC in ADNs with limited measurement deployments. 
Unlike previous works that utilize pseudo-measurement to improve the measurement condition, the proposed approach solves the problem of limited measurement deployments directly.
We first analyzed the problem of limited measurement deployments, which led to the problem of the partially observable state and unknown reward of DRL. Then, we propose the conservative critic and surrogate reward to address the two problems separately. 
The approach is adaptable to different limited measurement conditions of ADNs, and the minimum measurement condition is the active power of the root node and a few bus voltages, which has been fully demonstrated in the simulation results.
The effects of partial state, surrogate reward, and conservative critic are also analyzed in the simulation results.

Our proposed approach, designed for IB-VVC, may not be sufficient for ADNs with capacity banks, on-load tap changers, and storage devices. These tasks involve continuous and discrete actions.
To address this, in future works, we plan to extend our algorithm to handle mixed-integer optimization problems.



%

\printcredits

\bibliographystyle{elsarticle-num}

\bibliography{My_Library}

\begin{thebibliography}{10}
\expandafter\ifx\csname url\endcsname\relax
  \def\url#1{\texttt{#1}}\fi
\expandafter\ifx\csname urlprefix\endcsname\relax\def\urlprefix{URL }\fi
\expandafter\ifx\csname href\endcsname\relax
  \def\href#1#2{#2} \def\path#1{#1}\fi

\bibitem{farivarInverterVARControl2011}
M.~Farivar, C.~R. Clarke, S.~H. Low, K.~M. Chandy, Inverter var control for
  distribution systems with renewables, in: 2011 IEEE International Conference
  on Smart Grid Communications, 2011, pp. 457--462.

\bibitem{chenPhysicalassistedMultiagentGraph2023}
Y.~Chen, Y.~Liu, J.~Zhao, G.~Qiu, H.~Yin, Z.~Li, Physical-assisted multi-agent
  graph reinforcement learning enabled fast voltage regulation for pv-rich
  active distribution network, Applied Energy 351 (2023) 121743.

\bibitem{juBiLevelConsensusADMMBased2022}
Y.~Ju, Z.~Zhang, W.~Wu, W.~Liu, R.~Zhang, A bi-level consensus admm-based fully
  distributed inverter-based volt/var control method for active distribution
  networks, IEEE Transactions on Power Systems 37~(1) (2022) 476--487.

\bibitem{yanMultiAgentSafeGraph2024}
R.~Yan, Q.~Xing, Y.~Xu, Multi-agent safe graph reinforcement learning for pv
  inverters-based real-time decentralized volt/var control in zoned
  distribution networks, IEEE Transactions on Smart Grid 15~(1) (2024)
  299--311.

\bibitem{wangSafeOffPolicyDeep2020}
W.~Wang, N.~Yu, Y.~Gao, J.~Shi, Safe off-policy deep reinforcement learning
  algorithm for volt-var control in power distribution systems, IEEE
  Transactions on Smart Grid 11~(4) (2020) 3008--3018.

\bibitem{xuDatadrivenInverterbasedVolt2023}
T.~Xu, W.~Wu, Y.~Hong, J.~Yu, F.~Zhang, Data-driven inverter-based volt/var
  control for partially observable distribution networks, CSEE Journal of Power
  and Energy Systems 9~(2) (2023) 548--560.

\bibitem{caoPhysicsInformedGraphicalRepresentationEnabled2024}
D.~Cao, J.~Zhao, J.~Hu, Y.~Pei, Q.~Huang, Z.~Chen, W.~Hu, Physics-informed
  graphical representation-enabled deep reinforcement learning for robust
  distribution system voltage control, IEEE Transactions on Smart Grid 15~(1)
  (2024) 233--246.

\bibitem{azimianStateTopologyEstimation2022}
B.~Azimian, R.~S. Biswas, S.~Moshtagh, A.~Pal, L.~Tong, G.~Dasarathy, State and
  topology estimation for unobservable distribution systems using deep neural
  networks, IEEE Transactions on Instrumentation and Measurement 71 (2022)
  1--14.

\bibitem{primadiantoReviewDistributionSystem2017}
A.~Primadianto, C.-N. Lu, A review on distribution system state estimation,
  IEEE Transactions on Power Systems 32~(5) (2017) 3875--3883.

\bibitem{liuTwoStageDeepReinforcement2021}
H.~Liu, W.~Wu, Two-stage deep reinforcement learning for inverter-based
  volt-var control in active distribution networks, IEEE Transactions on Smart
  Grid 12~(3) (2021) 2037--2047.

\bibitem{liuRobustRegionalCoordination2021}
H.~Liu, C.~Zhang, Q.~Chai, K.~Meng, Q.~Guo, Z.~Y. Dong, Robust regional
  coordination of inverter-based volt/var control via multi-agent deep
  reinforcement learning, IEEE Transactions on Smart Grid 12~(6) (2021)
  5420--5433.

\bibitem{gaoModelaugmentedSafeReinforcement2022}
Y.~Gao, N.~Yu, Model-augmented safe reinforcement learning for volt-var control
  in power distribution networks, Applied Energy 313 (2022) 118762.

\bibitem{caoDeepReinforcementLearning2022}
D.~Cao, J.~Zhao, W.~Hu, N.~Yu, F.~Ding, Q.~Huang, Z.~Chen, Deep reinforcement
  learning enabled physical-model-free two-timescale voltage control method for
  active distribution systems, IEEE Transactions on Smart Grid 13~(1) (2022)
  149--165.

\bibitem{caoMultiAgentDeepReinforcement2020}
D.~Cao, W.~Hu, J.~Zhao, Q.~Huang, Z.~Chen, F.~Blaabjerg, A multi-agent deep
  reinforcement learning based voltage regulation using coordinated pv
  inverters, IEEE Transactions on Power Systems 35~(5) (2020) 4120--4123.

\bibitem{yiRealTimeSequentialSecurityConstrained2024}
Z.~Yi, X.~Wang, C.~Yang, C.~Yang, M.~Niu, W.~Yin, Real-time sequential
  security-constrained optimal power flow: A hybrid knowledge-data-driven
  reinforcement learning approach, IEEE Transactions on Power Systems 39~(1)
  (2024) 1664--1680.

\bibitem{sunOptimalVoltVar2024}
X.~Sun, Z.~Xu, J.~Qiu, H.~Liu, H.~Wu, Y.~Tao, Optimal volt/var control for
  unbalanced distribution networks with human-in-the-loop deep reinforcement
  learning, IEEE Transactions on Smart Grid 15~(3) (2024) 2639--2651.

\bibitem{peiMultiTaskReinforcementLearning2023}
Y.~Pei, J.~Zhao, Y.~Yao, F.~Ding, Multi-task reinforcement learning for
  distribution system voltage control with topology changes, IEEE Transactions
  on Smart Grid 14~(3) (2023) 2481--2484.

\bibitem{chenReinforcementLearningSelective2022}
X.~Chen, G.~Qu, Y.~Tang, S.~Low, N.~Li, Reinforcement learning for selective
  key applications in power systems: Recent advances and future challenges,
  IEEE Transactions on Smart Grid 13~(4) (2022) 2935--2958.

\bibitem{haarnojaSoftActorCriticAlgorithms2019}
T.~Haarnoja, A.~Zhou, K.~Hartikainen, G.~Tucker, S.~Ha, J.~Tan, V.~Kumar,
  H.~Zhu, A.~Gupta, P.~Abbeel, S.~Levine, Soft actor-critic algorithms and
  applications (Jan. 2019).
\newblock \href {http://arxiv.org/abs/1812.05905} {\path{arXiv:1812.05905}}.

\bibitem{schulmanProximalPolicyOptimization2017}
J.~Schulman, F.~Wolski, P.~Dhariwal, A.~Radford, O.~Klimov, Proximal policy
  optimization algorithms (Jul. 2017).
\newblock \href {http://arxiv.org/abs/1707.06347} {\path{arXiv:1707.06347}}.

\bibitem{mengPartialObservabilityDRL2022}
L.~Meng, R.~Gorbet, D.~Kuli{\'c}, Partial observability during drl for robot
  control (Sep. 2022).
\newblock \href {http://arxiv.org/abs/2209.04999} {\path{arXiv:2209.04999}}.

\bibitem{liuOnlineMultiAgentReinforcement2021}
H.~Liu, W.~Wu, Online multi-agent reinforcement learning for decentralized
  inverter-based volt-var control, IEEE Transactions on Smart Grid 12~(4)
  (2021) 2980--2990.

\bibitem{sunTwoStageVoltVar2021}
X.~Sun, J.~Qiu, Two-stage volt/var control in active distribution networks with
  multi-agent deep reinforcement learning method, IEEE Transactions on Smart
  Grid 12~(4) (2021) 2903--2912.

\bibitem{zhangHierarchicallyCoordinatedVoltageVAR2020}
C.~Zhang, Y.~Xu, Hierarchically-coordinated voltage/var control of distribution
  networks using pv inverters, IEEE Transactions on Smart Grid 11~(4) (2020)
  2942--2953.

\bibitem{zhangAdvancedElectricPower2010}
B.~Zhang, Z.~Yan, Advanced Electric Power Network Analysis, first edition
  Edition, Cengage Learning Asia, 2010.

\bibitem{caoDataDrivenMultiAgentDeep2021}
D.~Cao, J.~Zhao, W.~Hu, F.~Ding, Q.~Huang, Z.~Chen, F.~Blaabjerg, Data-driven
  multi-agent deep reinforcement learning for distribution system decentralized
  voltage control with high penetration of pvs, IEEE Transactions on Smart Grid
  12~(5) (2021) 4137--4150.

\bibitem{nguyenThreeStageInverterBasedPeak2022}
H.~T. Nguyen, D.-H. Choi, Three-stage inverter-based peak shaving and volt-var
  control in active distribution networks using online safe deep reinforcement
  learning, IEEE Transactions on Smart Grid 13~(4) (2022) 3266--3277.

\bibitem{koenkerQuantileRegression2005}
R.~Koenker, Quantile Regression, Cambridge University Press, Cambridge ; New
  York, 2005.

\bibitem{ghoshWhyGeneralizationRL2021}
D.~Ghosh, J.~Rahme, A.~Kumar, A.~Zhang, R.~P. Adams, S.~Levine, Why
  generalization in rl is difficult: Epistemic pomdps and implicit partial
  observability, in: Advances in Neural Information Processing Systems,
  Vol.~34, Curran Associates, Inc., 2021, pp. 25502--25515.

\bibitem{suttonReinforcementLearningSecond2018}
R.~S. Sutton, A.~G. Barto, Reinforcement Learning, Second Edition: An
  Introduction, MIT Press, 2018.

\bibitem{leiDataDrivenOptimalPower2021}
X.~Lei, Z.~Yang, J.~Yu, J.~Zhao, Q.~Gao, H.~Yu, Data-driven optimal power flow:
  A physics-informed machine learning approach, IEEE Transactions on Power
  Systems 36~(1) (2021) 346--354.

\bibitem{mestavBayesianStateEstimation2019}
K.~R. Mestav, J.~{Luengo-Rozas}, L.~Tong, Bayesian state estimation for
  unobservable distribution systems via deep learning, IEEE Transactions on
  Power Systems 34~(6) (2019) 4910--4920.
\newblock \href {http://arxiv.org/abs/1811.02756} {\path{arXiv:1811.02756}}.

\bibitem{dabneyDistributionalReinforcementLearning2018}
W.~Dabney, M.~Rowland, M.~G. Bellemare, R.~Munos, Distributional reinforcement
  learning with quantile regression, in: Proceedings of the Thirty-Second AAAI
  Conference on Artificial Intelligence, AAAI Press, New Orleans, Louisiana,
  USA, 2018, pp. 2892--2901.

\bibitem{liuTwoCriticDeepReinforcement2024}
Q.~Liu, Y.~Guo, L.~Deng, H.~Liu, D.~Li, H.~Sun, W.~Huang, Two-critic deep
  reinforcement learning for inverter-based volt-var control in active
  distribution networks, IEEE Transactions on Sustainable Energy (2024) 1--14.

\bibitem{zimmermanMATPOWERSteadyStateOperations2011}
R.~D. Zimmerman, C.~E. {Murillo-S{\'a}nchez}, R.~J. Thomas, Matpower:
  Steady-state operations, planning, and analysis tools for power systems
  research and education, IEEE Transactions on Power Systems 26~(1) (2011)
  12--19.

\bibitem{thurnerPandapowerOpenSourcePython2018}
L.~Thurner, A.~Scheidler, F.~Sch{\"a}fer, J.-H. Menke, J.~Dollichon, F.~Meier,
  S.~Meinecke, M.~Braun, Pandapower---an open-source python tool for convenient
  modeling, analysis, and optimization of electric power systems, IEEE
  Transactions on Power Systems 33~(6) (2018) 6510--6521.

\bibitem{qiongIkelqRDRL_PO2024}
L.~Qiong, \href{https://github.com/ikelq/RDRL\_PO}{Ikelq/rdrl\_po} (Apr. 2024).
\newline\urlprefix\url{https://github.com/ikelq/RDRL\_PO}

\end{thebibliography}



\end{document}